\documentclass[a4paper, 10pt, notitlepage]{article}
\usepackage[utf8]{inputenc}
\usepackage{graphicx}
\usepackage{caption}
\usepackage{subcaption}
\usepackage{color}
\usepackage{url}
\usepackage{geometry}

\usepackage[small]{titlesec} 
\usepackage{amssymb}
\usepackage{amsfonts}
\usepackage{amsthm}
\usepackage{empheq,amsmath}
\usepackage{gensymb}
\usepackage[switch]{lineno} 

\usepackage[table]{xcolor}
\usepackage{leftidx} 
\usepackage{bigints}

\usepackage[symbol]{footmisc}

\numberwithin{equation}{section} 

\theoremstyle{remark}

\renewenvironment{abstract}
{\begin{quote}
\noindent \rule{\linewidth}{.5pt}\par{\bfseries \abstractname.}}
{\medskip\noindent \rule{\linewidth}{.5pt}
\end{quote}
}

\begin{document}
\begin{center}
\large{\textbf{THE EVOLUTION OF A NON-AUTONOMOUS CHAOTIC SYSTEM UNDER NON-PERIODIC FORCING: A CLIMATE CHANGE EXAMPLE}}\\
\end{center}

\begin{center}
F. DE MELO VIR\'ISSIMO$^{1,}$\footnote{Corresponding Author. e-mail: f.de-melo-virissimo@lse.ac.uk.}, D. A. STAINFORTH$^{1,2}$, J. BR\"{O}CKER$^{3}$\\
\end{center}

\begin{center}
{\small \it $^1$Grantham Research Institute on Climate Change and the Environment, London School of Economics and Political Science, London, WC2A 2AE, United Kingdom}\\
{\small \it $^2$Department of Physics, University of Warwick, Coventry, United Kingdom}\\
{\small \it $^3$School of Mathematical Physical and Computational Sciences, University of Reading, Reading, RG6 6AX, United Kingdom}\\
\end{center}

\begin{center}
     \textbf{NOTE: THIS IS A NON-PEER REVIEWED PREPRINT SUBMITTED TO ARXIV}
\end{center}

{\small
\begin{abstract}
Complex Earth System Models are widely utilised to make conditional statements about the future climate under some assumptions about changes in future atmospheric greenhouse gas concentrations; these statements are often referred to as climate projections. The models themselves are high-dimensional nonlinear systems and it is common to discuss their behaviour in terms of attractors and low-dimensional nonlinear systems such as the canonical Lorenz `63 system. In a non-autonomous situation, for instance due to anthropogenic climate change, the relevant object is sometimes considered to be the pullback or snapshot attractor. The pullback attractor, however, is a collection of {\em all} plausible states of the system at a given time and therefore does not take into consideration our knowledge of the current state of the Earth System when making climate projections, and are therefore not very informative regarding annual to multi-decadal climate projections. In this article, we approach the problem of measuring and interpreting the mid-term climate of a model by using a low-dimensional, climate-like, nonlinear system with three timescales of variability, and non-periodic forcing. We introduce the concept of an {\em evolution set} which is dependent on the starting state of the system, and explore its links to different types of initial condition uncertainty and the rate of external forcing. We define the {\em convergence time} as the time that it takes for the distribution of one of the dependent variables to lose memory of its initial conditions. We suspect a connection between convergence times and the classical concept of mixing times but the precise nature of this connection needs to be explored. These results have implications for the design of influential climate and Earth System Model ensembles, and raise a number of issues of mathematical interest.
\\
\end{abstract}

\begin{center}
    \textbf{Keywords: non-autonomous systems, climate change, climate modelling, climate prediction, chaotic dynamical systems, pullback attractor}\\
\end{center}

\begin{center}
    \textbf{AMS classification: 37C60, 86A08 (Primary) 37N10 (Secondary)}
\end{center}
}

\newpage

\sloppy

\section{Introduction}\label{Sec1}

\medskip

The theory of non-autonomous nonlinear dynamical systems has enjoyed great popularity over the past few decades, particularly within the climate modelling community~\cite{ghil2019:nonlinearity}. This is because complex global climate models, or rather Earth System Models (ESMs), which are widely used to make projections of the 21st century and to support the IPCC's climate assessment reports, are subject to non-periodic, climate-change-like forcing, which inevitably breaks their autonomy. These models are also high-dimensional, multi-component, multi-scale, chaotic nonlinear systems and as a consequence, any forward computation - that is to say, projection of the future within the model - is highly sensitive to the finest details\footnote{Up to the precision of the computer.} of the initial state, making climate prediction a non-trivial task.

Uncertainty in the state from which to initialise ESMs is known as initial condition uncertainty (ICU). The sensitivity of such climate system models to ICU is well known since the early 60s'~\cite{lorenz63:chaos} and has led to the development of ensemble weather forecasting~\cite{toth2019:weatherforecast}. Its relevance for climate forecasting is also increasingly being recognised ~\cite{stainforth2007:royalsoc1,deser2012:internalvar,deser2012:comminternalvar,hawkins2016:neartermclimate}, as it is the necessity of using large climate initial condition ensembles (ICEs) to characterise ICU~\cite{daron2013:ensemble}. Nevertheless, it is often assumed that the uncertainty arising from ICU can be addressed by taking statistics from a single, long trajectory, which it is assumed, over time, would explore all possible states in phase space. In a stationary system\footnote{And periodically-forced ones as well, via a stroboscopic map.} this is essentially an ergodic~\cite{drotos2016:nonergodicity}, or ``kairodic''~\cite{daron2013:ensemble}, assumption: that averages and distributions of states over long periods (e.g. 30 years for IPCC) are representative of any particular instant - with the caveat that it would require infinite time for convergence. However, under non-periodic, climate-change-like forcing - such as increasing atmospheric greenhouse gas concentrations - the system is not ergodic, and hence cannot be studied in this way~\cite{drotos2016:nonergodicity}. 

The non-autonomous nature of ESMs under anthropogenic, non-periodic climate change forcing means that, in general, such a system do not possess an attractor in the classical sense, because we cannot take the asymptotic limit as time tends to infinity. Recent years have seen the emergence of a number of approaches from the mathematical community to address this issue~\cite{ghil2019:nonlinearity,tel2020:climrealisations}. Central to these approaches is the idea that a model's climate can be formally seen as an evolving probability distribution constructed from an ensemble of simulations which have been initialised from different ICs, initialised in the very remote past. This can be thought of as multiple ``evolutions'' of the same Earth System (that is to say, they all obey the same physical laws) but with each one starting from different initial points~\cite{tel2020:climrealisations}.

For a wide class of nonautonomous systems, it has been shown that, in this ``parallel climate realisations" approach, the correct concept to describe a time-dependent set in the phase space as the ``limit'' of a set of ICs is the {\it pullback attractor}.~\cite{cheban2002:pullback,kloeden2020:nonautonomous,ghil2008:climatedynamics,chekroun2011:pullback,pierini2016:doublegyre}. Many climate models (including the one discussed here\footnote{We are therefore confident that the existence of a pullback attractor can be established for the model used in this study but the proof will be left for future work.}) satisfy some form of energy balance which typically implies the core structural hypotheses required to establish the existence of pullback attractors. At any instant in time, the system's `climate' can therefore be taken as an instantaneous slice of the pullback attractor - this slice is the so-called {\it snapshot attractor}. Furthermore, in the same way that the (pullback) attractors are some form of ``limit'' for a set of IC's, the initial {\em distribution} of IC's might converge to a time dependent ``pullback'' probability measure supported on the pullback attractor. Invariant and pullback measures are typically not unique but here we are specifically interested in so--called {\em natural} or {\em  physical} pullback measures, which emerge as the limit of smooth IC distributions\footnote{We will not prove the existence of natural pullback measures for the system considered in this paper; rather, the importance of this concept is demonstrated, showing that such a proof would be worthwhile.}~\cite{newman2023:measures}. 

However, while mathematically appealing, these concepts are of limited use in supporting the construction of climate change ensembles of ESMs, and therefore in making climate projections and ultimately supporting society. By definition, the pullback attractor depends on initialisation infinitely far in the past\footnote{See Equation~(\ref{eq3.0-1}).}. Generally, this problem can be overcome by noting that in most cases we can assume that mixing happens on finite time scales, which, however long, can be taken as providing a convergence time: the time taken for the ensemble dynamics to forget its initial state. We do not therefore require infinitely long simulations, only sufficiently long, where "sufficient" is defined by this convergence time. Nevertheless this means that the pullback attractor is only applicable for long term climate analyses - longer than the convergence time. This convergence time can be small (around 5 years) for a simple conceptual low-dimensional atmospheric model system~\cite{drotos2015:snapshot1D} but rather long (over 150 years) even for fast-mixing atmosphere variables in an intermediate-complexity ESM~\cite{herein2016:snapshotGCM}. In other words, the pullback attractor approach might give us a good description of our idealised model system's climate by the end of the next century (i.e., in about 150 years time), but it can not tell us how we will get there.

This means that, while the pullback attractor represents the internal variability of the mathematical system on timescales beyond the convergence time, it is not the relevant object to represent climate on shorter timescales because it does not reflect knowledge regarding the current state of the climate system. On shorter timescales, the representative distribution is more constrained. The set of trajectories that make up this constrained distribution is a subset of those making up the pullback attractor, but it is not clear how the two distributions relate to each other.

Here we consider how to quantify this initial response and how such forward distributions can depend on both our knowledge of the current state and the characteristics of the non-autonomous forcing. These issues are critical to understanding what is required to make climate projections - even in the perfect model scenario~\cite{smith2002:pnasclimate} - and in characterising the behaviour of non-autonomous, non-periodic, nonlinear systems more broadly. To do so, we use a low-dimensional system with characteristics of an ESM~\cite{dmv2023:lowdimDS}. The concept of an {\em evolution set} is introduced to describe the set on which a more constrained distribution would be supported. We also introduce the concept of an {\em evolution distribution} to describe the more constrained distribution and we consider the convergence time for this evolution distribution to become indistinguishable from the pullback invariant distribution.

The paper is divided as follows. In Section 2, we describe the model used in this study, as well as the experiments performed. In Section 3, we elaborate on the concept of the pullback attractor, demonstrate it with examples from our model, and define and illustrate the convergence time for different variables in a stationary situation. In Section 4, we approach the transient climate change problem in combination with some hypothetical, highly-constrained knowledge of the initial state - so-called micro ICUs~\cite{stainforth2007:royalsoc1,hawkins2016:neartermclimate}. In Section 5 we consider situations where the initial state is not well constrained - so called macro ICUs~\cite{stainforth2007:royalsoc1,hawkins2016:neartermclimate}, while revisiting the concept of convergence time in the non-autonomous situation. In Section 6, we explore the influence of the forcing on the evolving distributions. We then conclude the paper with Section 7, where we discuss further questions and future directions for the this study.

\section{Modelling framework}\label{Sec2}

\subsection{Model}\label{Sec2.1}
We use a low-dimensional coupled ocean-atmosphere model, which is taken as a conceptual representation of a climate model. In this model, the ocean domain is presented as two connected but distinct basins, say, one representing the ocean at high latitudes and another representing it at low latitudes in the same hemisphere, with its dynamics given by the Stommel `61 (S61) model~\cite{stommel61:S61model}. The S61 model is based on the free convection controlled by density differences maintained by heat and salt exchange between the reservoirs. The atmosphere is represented by a simplified description of its large scale circulation in one hemisphere, given by the Lorenz `84 (L84) model~\cite{lorenz1984:L84model,lorenz1990:L84modelintransivity}. The L84 model is based on the interaction of the westerly, mid-latitude wind current and large scale, pole-ward eddies.

The L84 model and the S61 model form the coupled ocean-atmosphere model used in this study, which we shall refer to as Lorenz 84-Stommel 61 (L84-S61) model. 

Mathematically, the L84-S61 model consists on the following five coupled ODEs
\begin{flalign}
    &X' = -Y^{2} - Z^{2} -aX + a(F_{0}(t) + F_{1}T) \label{eq2.1-1}\\
    &Y' = XY - bXZ - Y + G_{0} + G_{1}(T_{av}-T) \label{eq2.1-2}\\
    &Z' = bXY + XZ - Z \label{eq2.1-3}\\
    &T' = k_{a}(\gamma X - T) - |f(T,S)|T - k_{w}T \label{eq2.1-4}\\
    &S' = \delta_{0} + \delta_{1}(Y^{2}+Z^{2}) - |f(T,S)|S - k_{w}S \label{eq2.1-5}
\end{flalign}
where
\begin{flalign}
    & f(T,S) = \omega T - \epsilon S \label{eq2.1-6}\\
    & F_{0}(t) = F_{m} + M \cos((2\pi t/K) - \pi/12) + F_{\mathrm{CC}}(t) \label{eq2.1-7}
\end{flalign}
and
\begin{equation} 
  F_{\mathrm{CC}}(t) = \begin{cases}
 0  & \text{if $t < t_{\mathrm{start}}$} \\
 (H/K)(t-t_{\mathrm{start}}) & \text{if $t_{\mathrm{start}} \leq t \leq t_{\mathrm{end}}$} \\
 (H/K)(t_{\mathrm{end}}-t_{\mathrm{start}}) & \text{if $t_{\mathrm{end}} < t$}.
\end{cases}
\label{eq2.1-8} 
\end{equation}
In the above, the variables $X,Y,Z$ represent the high-frequency, atmospheric variables from the L84 model: $X$ represents the intensity of the symmetric westerly wind, $Y$ and $Z$ are the Fourier amplitudes characterising a chain of large-scale eddies, which transport heat towards the pole at a rate proportional to their amplitude. The variables $T,S$ are the slow ocean variables as in the S61 model: $T$ and $S$ denote the pole-equator temperature and salinity differences, respectively.
The function $f(T,S)$ represents the strength if the thermohaline circulation (THC), while $F_{0}(t)$ is the forcing due to seasonal variation in the heating contrast between the pole and equator. The latter corresponds to an average forcing equals $F_{m}$ which varies seasonally according to a cosine function with amplitude $M$, and can be forced towards another value at a rate $H$. All the variables in the model are non-dimensional. The model parameters and their reference values are described in Table S.1, except the forcing function $F_{0}(t)$ which are presented separately in Table~\ref{table1}.

While $t$ denotes the non-dimensional time, we note that the characteristic time for the this model is 5 days, and hence, one time unit in this model corresponds to 5 days, as originally assumed by Lorenz (1984)~\cite{lorenz1984:L84model}. We refer to this as 1 Lorenz Time Unit (LTU). Hence, a 365-day year has $K=73$ LTUs.

\begin{table}[h!]
	\centering
    \begin{tabular}{ c | c | p{10cm} }
    \hline
    \hline
    \textbf{Parameter} & \textbf{Value} & \textbf{Description} \\ \hline
    $F_{m}$ & 7 & 1-year mean value of the seasonal variation function $F_{0}(t)$ when $H=0$ \\ 
    $H$ & 0.01 & Externally forced rate of change of $F_{m}$  \\ 
    $M$ & 1 & Magnitude of the seasonal cycle  \\ 
    $(t_{\mathrm{start}}/K)$ & 0 & Start of non-periodic external forcing (in years)  \\ 
    $(t_{\mathrm{end}}/K)$ & 100 & End of non-periodic external forcing (in years) \\ \hline \hline
    \end{tabular}
  \caption{Description of the parameters and their reference values in the forcing function $F_{0}(t)$, as per Daron and Stainforth (2013)~\cite{daron2013:ensemble}.}
    \label{table1}
\end{table}

The L84-S61 model is a nonlinear, non-autonomous system of ODEs~\cite{sell1967:nonautonomous}. Using vector notation, this system can be written as
\begin{equation}
    \mathbf{X}' = \mathbf{F}(\mathbf{X},t) \label{eq2.1-9},
\end{equation}
where $\mathbf{X}=(X,Y,Z,T,S)$, and $\mathbf{F}(\mathbf{X},t)$ is a time-dependent, nonlinear vector function of $\mathbf{X}$ given by the right-hand side of Equations~(\ref{eq2.1-1}) to~(\ref{eq2.1-5}). Its solutions are bounded, i.e. $||\mathbf{X}|| < C$, with $C$ being a positive constant. The system is conditionally dissipative, i.e. $\nabla \cdot \mathbf{F}(\mathbf{X},t) < 0$ under certain conditions, meaning that finite-volume attractors might exist.

Despite being a simplified representation of the ocean-atmosphere system, the L84-S61 model retain some of main characteristics of a state-of-the-art ESM: it is nonlinear, multiscale, multi-component, complex and chaotic. Hence, conceptual results obtained from this model can be insightful, if not informative, of general properties of ESMs. However, contrary to complex ESMs, which are high-dimensional (normally with billions or even trillions of degrees of freedom), the L84-S61 model consists of only 5 ODEs, making it an affordable model to be (extensively) studied computationally - in particular, allowing for very large ensembles to be run.

The L84-S61 model was first derived by Van Veen et al. (2001)~\cite{vanveen2001:L84S61}, with a similar model appearing in Roebber (1995)~\cite{roebber1995:L84S61}. The version presented here is the same used in Daron and Stainforth (2013)~\cite{daron2013:ensemble}. For details on the derivation of the L84-S61 model, the reader is suggested to consult Van Veen et al. (2001)~\cite{vanveen2001:L84S61}. Details on the individual model components can be found on the original works of Stommel (1961)~\cite{stommel61:S61model} and Lorenz (1984, 1990)~~\cite{lorenz1984:L84model,lorenz1990:L84modelintransivity}. A didactic introduction to the L84 model can also be found in Provenzale and Balmforth (1999)~\cite{provenzale1999:chaos}.

\subsection{Numerical solver, parameter values and ensemble design}\label{Sec2.2}

The L84-S61 model is solved using the 4th-order Runge-Kutta method, with time step 0.01 LTUs (1.2 hours). The output frequency is 0.2 LTUs (1 day). All results, whether single trajectories or ensembles, are presented as 1-year averages\footnote{Strictu sensu, averages are not a solution to the L84-S61 system's IVP. However, for the concepts and computational results presented in this paper, this difference is of little importance. In fact, for temperature and salinity, the difference between annual averages and actual values is small, and hence the latter can be used instead as proof of concept. For the atmosphere, it only matters if we were to look at observables where the annual average of the observable is very different from the observable of the annual average. Such a function would have to be nonlinear to begin with. The only point where the difference potentially matters is in the convergence time for the atmosphere.}.

All simulations use the parameter values as shown in Tables~\ref{table1} and S.1, except for some simulations in Sections~\ref{Sec3} and~\ref{Sec6}, in which $H=0$ and $0.0025$, respectively. Regarding the forcing, note that the values presented in Table~\ref{table1} means that the forcing oscillates seasonally around an average value $F_m=7$ with seasonal amplitude $M=1$, while being driving to another value at a rate of $H=0.01$ units per year, or 1 unit per 100 years.

The ensembles run in this work are designed as follows. Given an initial condition $\mathbf{X}_{0}=(X_{0,1},X_{0,2},X_{0,3},X_{0,4},X_{0,5})$ in the phase space, we randomly sample another 1,000 initial conditions such that, for each dependent variable, the sample is normally distributed around $X_{0,j}$ with variance given by $\sigma_{X_{0,j}}$ - with $\sigma_{X_{0,j}}$ being two orders of magnitude lower than $X_{0,j}$. Hence, each ensemble has 1,001 members. The details of each individual experiment, including duration and parameter values, can be found in the Supplementary Materials.

\section{The pullback attractor and convergence time}\label{Sec3}

The pullback attractor~\cite{cheban2002:pullback,kloeden2020:nonautonomous} is a mathematical object that generalises the concept of attractor to non-autonomous dynamical systems. This approach consists on the idea that, for most non-autonomous systems, there exists a time-dependent object in the phase space, to which trajectories that started in the infinite past will converge. Such object presents therefore a natural distribution for the internal variability of the system.

A formal definition can be presented as follows. Let us denote the solution to the initial value problem (IVP) given by Equation~(\ref{eq2.1-9}) and $\mathbf{X}(t_{0})=\mathbf{X}_{0}$ by $\mathbf{X}(t;\mathbf{X}_{0},t_{0})$, and the corresponding phase space by $\mathbb{X}$. A set $\mathcal{A}=\mathcal{A}(t)$ in the phase space is said to ``pullback'' attract a set, or ensemble of points $D_{\mathbf{X}_{0}} \subseteq \mathbb{X}$ if, for all $\mathbf{Y} \in D_{\mathbf{X}_{0}}$,

\begin{equation}
    \text{dist}_{\mathbb{X}}\left(\mathbf{X}(t;\mathbf{Y},t_{0}),\mathcal{A}(t)\right) \longrightarrow 0 \text{ as } t_{0} \longrightarrow -\infty, \label{eq3.0-1}
\end{equation}
for all $t$, where $\text{dist}_{\mathbb{X}}(\cdot,\cdot)$ denotes the Hausdorff semi-distance between sets in the phase space. The time-dependent set $\mathcal{A}(t)$, if also invariant with respect to the dynamics, is called {\em pullback attractor}. When pullback attractors exist, there might also exists an invariant probability distribution supported on this set, so-called the {\em pullback invariant measure} (or {\em distribution}), which we will generically denote by $\mu_{\mathcal{A}}$~\cite{chekroun2011:pullback}.

An explicit, rigorous computation of both $\mathcal{A}(t)$ and $\mu_{\mathcal{A}}$ is only viable for very simple dynamical systems, and usually not possible for most nonlinear ones, including L84-S61. However, for non-conservative systems, a more practical approach is possible. This relies on the fact that, in general, a solution (or ensemble) starting near or on the attractor takes only a finite time to lose most of its dependency on the initial condition and run through (span) most of the attractor. The time for this convergence is dependent on the system and its relevant time scales, and can also be estimated numerically, as we shall see below.

Figures~\ref{fig1}(a-c) illustrate this convergence to the pullback attractor for some of the variables of L84-S61. There, the pullback attractor and its natural distribution are computed from a micro ICE normally distributed around a central IC point $\mathbf{X}_{0}$ in the attractor, with variance $\sigma_{\mathbf{X}_{0}}$ being $\mathcal{O}(10^{-2})$ for atmosphere variables, $\mathcal{O}(10^{-3})$ for the ocean temperature and $\mathcal{O}(10^{-4})$ for ocean salinity (as per Daron and Stainforth, 2013~\cite{daron2013:ensemble}; see also Supplementary Materials). Note that, soon after the simulation starts, the initial micro cluster of trajectories disperses quickly and cover most of the attractor within a few years. The exact number of years depends on the variable of consideration though. For example, the time taken is visibly long for the ocean temperature (Figure~\ref{fig1}(a)), and even longer for the salinity (Figure~\ref{fig1}(b)), but very short for the fast, atmospheric variable $X$ (Figure~\ref{fig1}(c)). The latter is in line with what has been reported by Drot\'os et al. (2015)~\cite{drotos2015:snapshot1D} and T\'el et al. (2020)~\cite{tel2020:climrealisations} for the L84 atmospheric model.

\begin{figure}
  \centering
\includegraphics[width=1\textwidth]{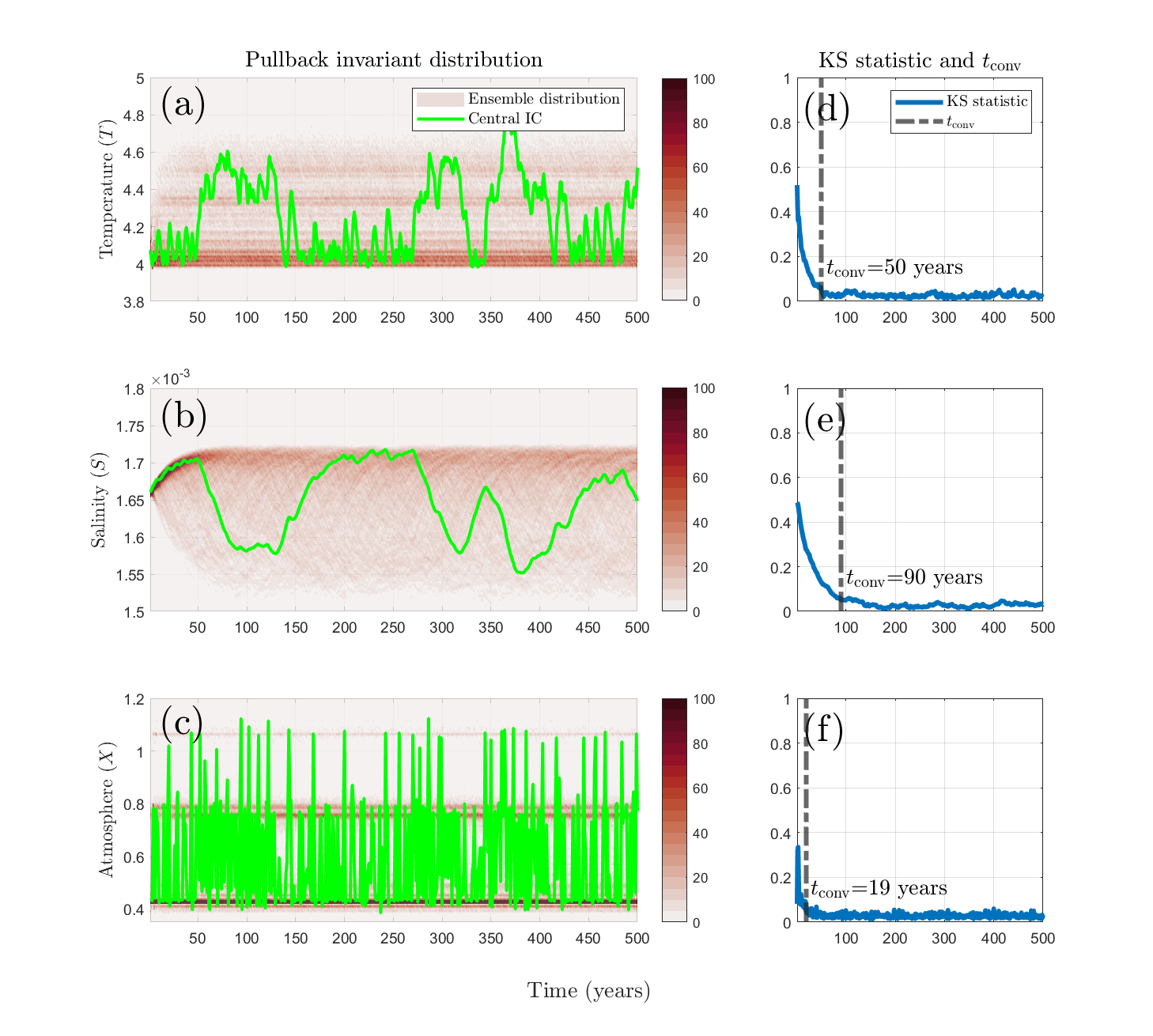}
\caption{Left column: Pullback attractor and its natural distribution for L84-S61 computed from a 500 years micro ICE simulation, where green solid line shows the numerical solution starting from the central IC. Right column: Corresponding convergence time computed using Equation~(\ref{eq3.1-1}) and the KS statistic based on a 100,000 single trajectory simulation. (a,d) ocean temperature; (b,e) ocean salinity; (c,f) atmosphere variable $X$ (intensity of westerly wind).}
\label{fig1}
\end{figure}

\subsection{Convergence time}\label{Sec3.1}

The convergence time, which we shall denote as $t_{\mathrm{conv}}$, can be loosely defined as the time taken by a localised ensemble to become indistinguishable from the pullback attractor. A statistically formal way to compute $t_{\mathrm{conv}}$ is by comparing, at each instant of time, the distribution of interest with a snapshot of the numerically estimated pullback invariant distribution, via a hypothesis test using some suitable statistics, where the null hypothesis $H_{0}$ is that both distributions come from the same population. If we define a function of time $h$ such that $h(t) = 1$ if the null hypothesis is rejected at time $t$ and $h(t) = 0$ if not rejected, then we could define $t_{\mathrm{conv}}$ such that

\begin{gather}
    t_{\mathrm{conv}} = \inf\{(t/K) \in [(t_{0}/K), \infty): h(t) = 0\}. \label{eq3.1-1}
\end{gather}

In the definition above\footnote{Note that we opted to define $t_{\mathrm{conv}}$ as normalised by $K$, so that the corresponding unit is year, instead of LTU.}, there might exist $t>t_{\mathrm{conv}}K$ such that $h(t) = 1$, which might put in question whether the convergence has been achieved. To avoid that, a statistically robust way to define $t_{\mathrm{conv}}$ would be to take the distribution of $h(t)$ in the time interval of consideration, repeat the experiment several times, and build the distribution of $h(t)$ values for all those experiments, which can then be translated into a distribution of $t$ values, with associated uncertainties. This resulting distribution should cluster around a value $\overline{t}$ that would be taken as $t_{\mathrm{conv}}$.

Both ways of estimating $t_{\mathrm{conv}}$ are clearly dependent on the system of interest, as well as the initial condition and also the dynamical variable in question. Crucially, in practice, when dealing with computationally-generated distributions, such computation is also dependent on the size of the ensemble. There is not a unique way of doing it, and hence $t_{\mathrm{conv}}$ is also dependent on the test used, as well as the significance level chosen. There are several ways to test this hypothesis~\cite{stephens1974:EDFstats}. In this work, we use a two-sample Kolmogorov-Smirnov (KS) test~\cite{massey1951:KSstats}. For two distributions $P_{1,n_{1}}(x)$ and $P_{2,n_{2}}(x)$ of sizes $n_{1}$ and $n_{2}$ respectively, the KS test is defined as

\begin{equation}
    D(P_{1,n_{1}},P_{2,n_{2}}) = \sup\limits_{x} |P_{1,n_{1}}(x) - P_{2,n_{2}}(x)|. \label{eq3.1-2}
\end{equation}
For the KS test, null hypothesis $H_{0}$ should be rejected\footnote{For convenience, in this work we use MATLAB's build-in function \texttt{kstest2} instead. This function rejects the null hypothesis based on the p-value, and not by comparing the test statistic with a reference value.} at significance at level $\alpha$ if $D(P_{1,n_{1}},P_{2,n_{2}}) > C_{n_{1},n_{2},1-\alpha}$, where $C_{n_{1},n_{2},1-\alpha}$ can be found in~\cite{book2008:kstest}.  

We illustrate this approach by computing $t_{\mathrm{conv}}$ for the spinup distribution shown in Figures~\ref{fig1}(a-c). To do so, we test $H_{0}$ with significance level $\alpha = 0.05$, where the reference distribution is given by a 100,000 years single-trajectory solution starting from the same central IC (Supplementary Materials). This is presented in Figures~\ref{fig1}(d-f), which shows that $t_{\mathrm{conv}}$ is 90 years for salinity, 50 years for temperature, but only 19 years for atmosphere. The latter is substantially higher than what has been reported by Drot\'os et al. (2015)~\cite{drotos2015:snapshot1D}, which found a $t_{\mathrm{conv}}$ of only 5 years for the L84 system, suggesting that the coupling with slow-mixing variables increases the relaxation period for the atmosphere variables in this context.

An alternative way to define a convergence time would be to assume that initially the statistic - in this case the KS statistic - $D$ decays exponentially, such that $D(t) \approx D(t_{0})\exp^{-\tau (t-t_{0})}$. In this case, such a convergence time could be taken as $1/\tau$. The characteristic decay exponent can be estimated by looking at the logarithm of $D$, which is presented in Figure S.1, and computing the angular coefficient of the straight line it approaches in the first few years of decay. This gives $\tau$ equals 0.0378, 0.264 and 0.1221 for $T$, $S$ and $X$ respectively. These correspond to estimated times of approximately 26 years, 38 years and 8 years respectively, which is roughly half the values of $t_{\mathrm{conv}}$ estimated via Equation~(\ref{eq3.1-1}). Hence although quantitatively different, both approaches provide very similar information.

\subsection{Caveats with the pullback attractor approach}\label{Sec3.2}

The pullback approach has been proposed as an alternative way of defining climate: it gives a mathematically sound measure of the system's internal variability, and being time dependent, provides both a natural set of plausible states at each instant of time - the snapshot attractor - and a
natural probability distribution of events at each instant of time - the pullback invariant distribution. This has been discussed and illustrated by several authors~\cite{ghil2019:nonlinearity,kovacs2020:epidemics}, and has proven to be a more rigorous and useful definition of climate for long-term (e.g. IPCC-like) future scenarios. 

This approach comes with some caveats though. By definition, the computation of such object requires an ensemble to be initialised in the infinite past, which is impractical from the computational point of view. In general, it is possible to approximately compute the attractor provided that the system is run for longer than $t_{\mathrm{conv}}$. But again, this is problematic, particularly in climate modelling: on one hand, some components of the Earth System evolve on long timescales of hundreds to thousands of years; on the other hand, anthropogenic, non-periodic forcing started only a couple of centuries ago.

Another caveat is that, while the pullback attractor represents all the internal variability of the mathematical model, it is known that only a few of these states can be representative of today's climate. Therefore, using the pullback attractor to measure ``tomorrow's" climate might include a large number of unrealistic states - they are part of the internal dynamics of the model but not attainable within that time frame for a given initial condition. This will be discussed in the next section.

\section{Micro initial condition ensembles and the evolution set}\label{Sec4}

Although the pullback attractor provides a useful, mathematically sound definition for long-term climate (beyond the convergence time), it is less useful in quantifying the variability in the short-mid term (months to years, or even decades), when the intermittency of the dynamics is still dependent on the initial state of the system. This is because it overestimate the forecast uncertainty by allowing all possible states within the attractor, including those that do not reflect our knowledge of the present state of the system. 

For example, considering the snapshot attractor for a given day (say ``today''), it corresponds to a large range of possible values. But given sufficient information, it might be that only one of those states is possible (up to a certain level of residual uncertainty), so many of the states on the snapshot attractor are unrealistic
given our knowledge of ``today's'' system. We also know that the climate today constrains the climate of tomorrow, in the range whose the system still carries the memory from the initial state - that therefore excludes a large portion of the pullback attractor. This means that any snapshots of the pullback attractor over-quantifies the variability and distorts the probability of events in the short and mid-term.

\begin{figure}
  \centering
\includegraphics[width=1\textwidth]{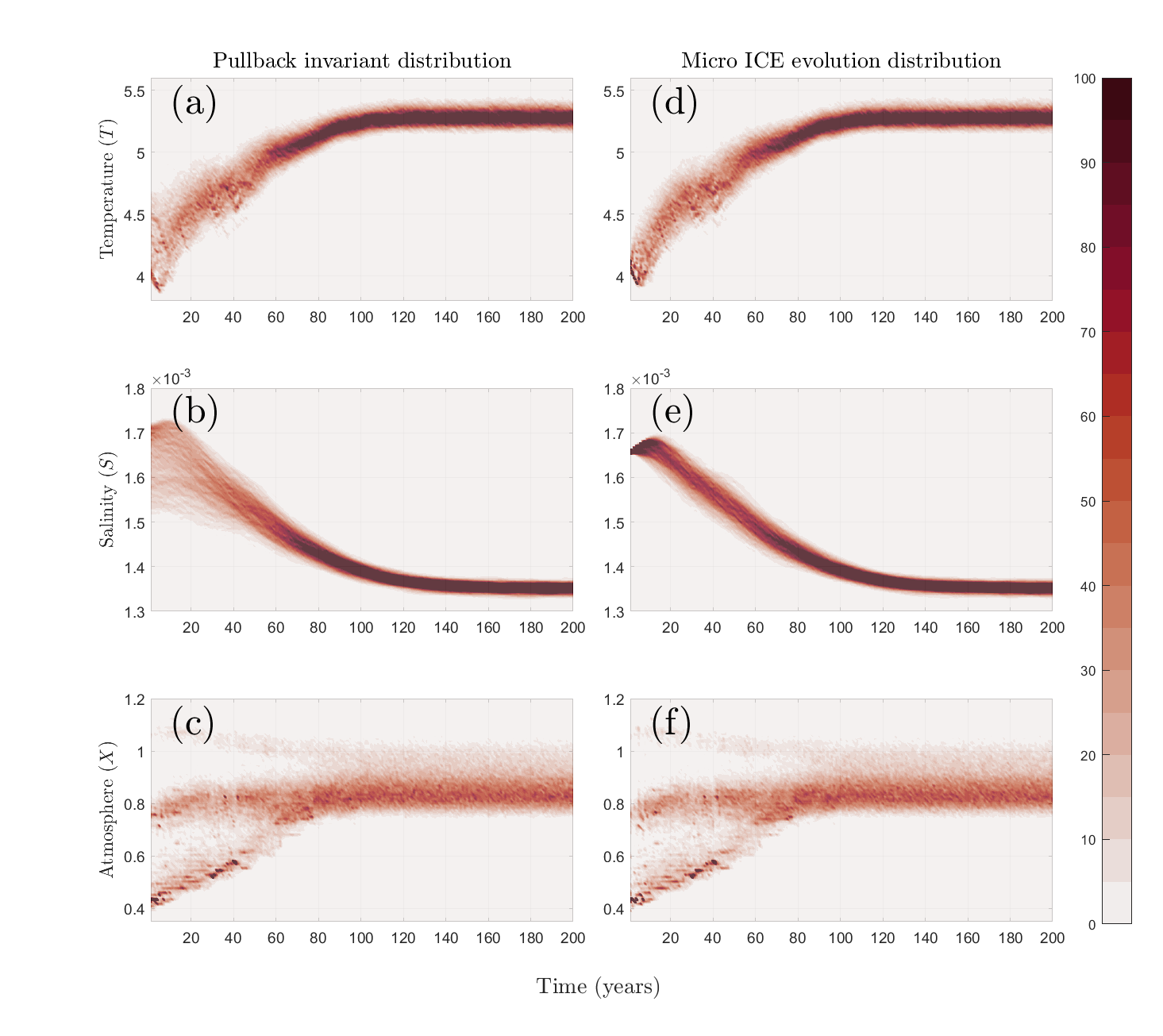}
\caption{Comparing the pullback invariant distribution with the distribution generated by a micro ICE, with $H=0.01$ in the first 100 years, and $H=0$ in the remaining 100 years. Left column shows the evolution of an ensemble which initially covers the entire pullback attractor. Right column shows the evolution of a micro ICE. Panels (a-f) show: (a,d) ocean temperature; (b,e) ocean salinity; (c,f) atmosphere variable $X$ (intensity of westerly wind).}
\label{fig2}
\end{figure}

This is illustrated in Figure~\ref{fig2}, where we present the evolution of a micro ICE under climate change next to the evolution of the pullback ICE of Figure~\ref{fig1}. This side-by-side comparison (see also Figure S.2 in the Supplementary Materials) shows that, in the first few decades, the pullback natural distribution, which is intrinsic to the mathematical system, over-represents climate uncertainty. Note that the evolution of the micro ICE is initially constrained to an smaller set, which is evolving over time, and seems to converge to the pullback attractor $\mathcal{A}(t)$ only after a few decades. For this reason, we name this as the {\em evolution set} $\mathcal{E}(t)$.

For a given non-autonomous chaotic system, this set is solely dependent\footnote{In practice, any numerical estimate of $\mathcal{E}$ will also depend on the size and shape of the initial ensemble.} on the initial state $\mathbf{X}_{0}$, on the initial micro-uncertainty given by the variance $\sigma_{\mathbf{X}_{0}}$ and on the initial time $t_{0}$. Therefore, we shall denote the evolution set as $\mathcal{E}(t;\mathbf{X}_{0},t_{0},\sigma_{\mathbf{X}_{0}})$. Some basic properties of this set are straightfowward. First, its existence is guaranteed by the existence and uniqueness for the IVP for ~(\ref{eq2.1-9}). Second, by the definition, we have that $\mathcal{E}(t_{0}) = \mathcal{D}_{\mathbf{X}_{0}}$, the ICE set. Also by definition, we have that $\mathcal{E}(t;\mathbf{X}_{0},t_{0},\sigma_{\mathbf{X}_{0}}) \longrightarrow \mathcal{A}(t)$ as $t_{0} \longrightarrow -\infty$. It also follows that, for an initial ensemble set within the pullback attractor, i.e. $D_{\mathbf{X}_{0}} \subseteq \mathcal{A}(t_{0})$, we have that $\mathcal{E}(t) \subseteq \mathcal{A}(t)$, for all $t\geq t_{0}$. In practice, when estimating both $\mathcal{E}(t)$ and $\mathcal{A}(t)$ numerically, these properties do not hold {\it ipsis literis}, and the design of the ensemble becomes quite important. We also note that, associated to $\mathcal{E}(t)$, this numerical example suggests the existence of a distribution $\mu_{\mathcal{E}}$ supported on this set, which we will assume to be true. Its relationship to the pullback invariant distribution $\mu_{\mathcal{A}}$ is not as clear though.

Climate modellers are familiar with the idea of exploring ICU using micro ICEs. Nevertheless, they are in general taken simply as an exploration of uncertainty, rather than the object we are trying to characterise. Here, we bring together the ideas of the pullback attractor with the methods applied in climate modelling and produce an attractor-like object which essentially represents future climate under climate change - which we called the evolution set.

The formalism above allow us to revisit the content of previous section, and reframe it in terms of ``forward'' convergence\footnote{Here, we note again that, pullback attractors are not forward attractors in general. Although, under certain conditions, a pullback attractor could satisfy a weak form of forward convergence. The interested reader can find a detailed exposition of this in the section 9.5 of Kloeden and Yang (2020)~\cite{kloeden2020:nonautonomous}.}. There, the existence of a convergence time $t_{\mathrm{conv}}$ might suggest that $\mathcal{E}(t) \approx \mathcal{A}(t)$ almost everywhere for $t>t_{\mathrm{conv}}$. Hence, the question is: does that really happen? Which conditions are necessary to prove that, for $t>>t_{0}$: 1) $\mathcal{E}(t)$ and $\mathcal{A}(t)$ are sufficiently close\footnote{Note that two sets might be infinitely close but disjoint. For instance, the sets of rational and irrational points within the interval $[0,1]$ are disjoint but their closure (with respect to the standard topology) equals the full interval.}; 2) $\mu_{\mathcal{A}}$ approximates $\mu_{\mathcal{E}}$ as $t \longrightarrow t_{\mathrm{conv}}$. If such statements are true, the pair $(\mathcal{A},\mu_{\mathcal{A}})$ would hold key mathematical information regarding future climate.

In the next section, we will explore some features of the evolution set by looking at its dependence on the initial conditions.

\section{Macro initial condition uncertainty}\label{Sec5}

Another issue related to the short-to-mid term climate prediction is the level of uncertainty of the actual state of the system in some variables. While small uncertainty can be covered by a micro IC ensemble, the uncertainty in the initial state of some variables might be of the same order of magnitude of the typical values for the variable itself, for instance if the initial state is based on a model spinup, or derived from the interpolation of sparse datasets, or even because of a lack of data.

From a climate prediction point of view, these are relevant, and macroscale variations in ocean quantities such as temperature and salinity, and atmospheric ICU have already been linked to decadal variations in regional climate in the Northern Hemisphere~\cite{hawkins2016:neartermclimate}. The question is therefore how would such macro uncertainty impacts the evolution of the system, via its evolution $\mathcal{E}(t)$ set. 

\subsection{Macro ICU from a control simulation (single trajectory)}\label{Sec5.1}

\begin{figure}
  \centering
\includegraphics[width=1\textwidth]{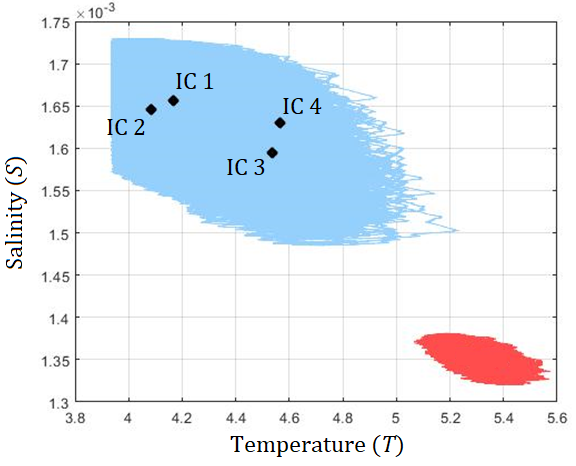}
\caption{Attractor for the system L84-S61 with $H=0$ when $F_m=7$ (blue) and $F_m=8$ (red) projected on the ocean temperature-salinity $(T,S)$ subspace. The black dots on the $F_m=7$ attractor indicated the location of ICs 1 to 4.}
\label{fig3}
\end{figure}

One of the sources of macro ICU is the potential to initiate climate ensembles from different states - including ocean states - from a long control run with an ESM. To illustrate this, we chose four different points in the attractor, all corresponding to a point in an existing trajectory after an initial 5,000 years spinup. For simplicity, we name these ICs by IC 1, IC 2, IC 3 and IC 4, with corresponding micro ICEs referred as ICE 1, ICE 2 and so on. Note that those ICs differ in all five dependent variables, and are illustrated in Figure~\ref{fig3} for the ocean variables. All ensemble distributions have the same variance, as noted in Section~\ref{Sec2.2}.

\begin{figure}
  \centering
\includegraphics[width=1\textwidth]{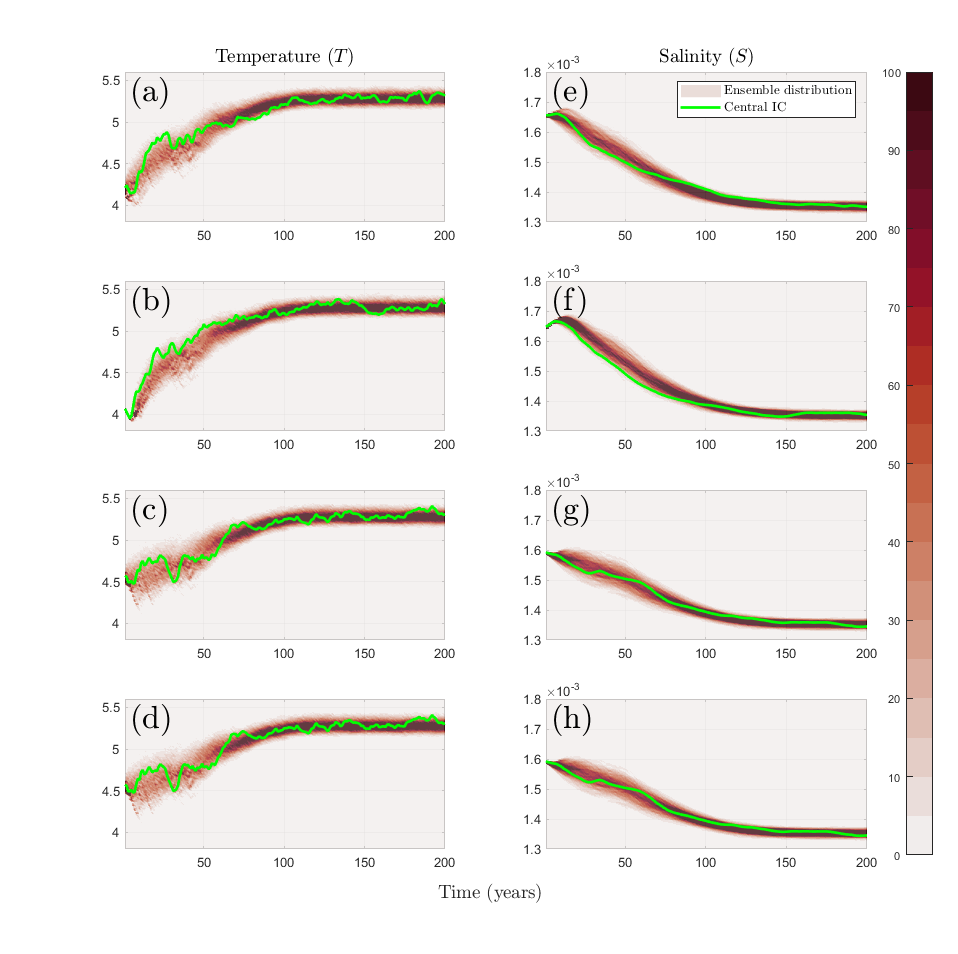}
\caption{Macro ICU from a control run simulation: comparing the evolution set and distribution of the slow-mixing ocean variables for different micro ICEs in a macro ICU scenario, with $H=0.01$ in the first 100 years, and $H=0$ in the remaining 100 years. Left column shows the ocean temperature. Right column shows ocean salinity. Panels (a-f) show: (a,e) IC 1; (b,f) IC 2; (c,g) IC 3; and (d,h) IC 4.}
\label{fig4}
\end{figure}

Figure~\ref{fig4} shows that, for the ocean variables, the dependence on the initial condition is significant. A first remark is that all four micro-initialised resulting distributions differ substantially from the pullback invariant distributions shown in Figures~\ref{fig2}(a,b). Further to that, they are also different among themselves. For instance, in Figure~\ref{fig4}(b), the micro ICE centred at IC 2 starting from a low temperature tends to decrease for a few years before increasing again, despite the monotonic increase in forcing. This is not followed by the micro ICE centred at the nearby IC 1, as shown in Figure~\ref{fig4}(a) which spreads out very quickly after initialisation and is visually (increasing) monotonic from the beginning. In the case of IC 2, the decrease in temperature is accompanied by an initial increase in salinity as shown in Figure~\ref{fig4}(f), which is then followed by a steady decrease.
Nevertheless, in all four cases, the distributions seems to coincide after a few decades, becoming visually indistinguishable from each other and from the pullback invariant distribution (see also Figure S.3).

This macro ICU dependence has important consequences for climate prediction in seasonal to decadal time scales. A common practice in climate modelling is to start a simulation from initial conditions obtained from a spinup ``control'' run. This control run allows one to find the system's attractor, but does not resolve the uncertainty about where in the attractor one should start from. As we have seen, different micro ICEs could lead to different transient distributions, representing a different climate in the short-to-mid term - even if the initial condition is obtained from the same solution after spinup.

\subsection{Macro ICU that reflect uncertainty in one variable}\label{Sec5.2}

Another source of macro ICU is when initialising the model from observations, in which case the uncertainty in some variables could be orders of magnitude higher than others. As an example, if in-situ data is being used to initialise the model, it is possible that one might have measurements for one variable but not for others, for instance in case of defective equipment (e.g. via bio-fouling). In this case, the initial state of the variable is subjected to macro uncertainty. 

\begin{figure}
  \centering
\includegraphics[width=1\textwidth]{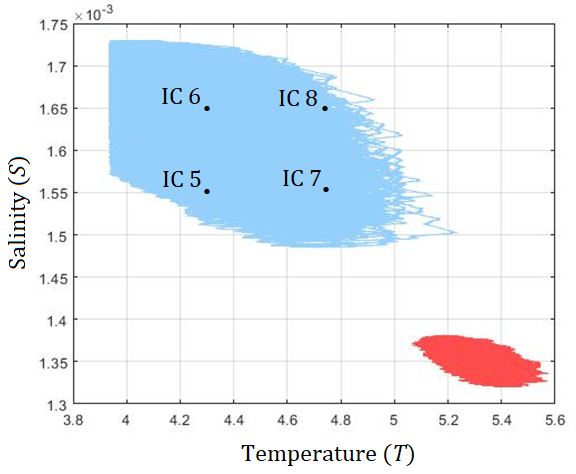}
\caption{Attractor for the system L84-S61 with $H=0$ when $F_m=7$ (blue) and $F_m=8$ (red) projected on the ocean temperature-salinity $(T,S)$ subspace. The black dots on the $F_m=7$ attractor indicated the location of ICs 5 to 8.}
\label{fig5}
\end{figure}

This scenario is illustrated in Figure~\ref{fig5}, where we highlighted four possible initial conditions, named by IC 5, IC 6, IC 7 and I 8, which are identical in the atmosphere variables $X,Y,Z$, but may differ in temperature and salinity (Supplementary Materials). For instance, IC 5 and IC 6 has identical temperature but differ in salinity; the converse is true for IC 6 and IC 7, and so on.

\begin{figure}
  \centering
\includegraphics[width=1\textwidth]{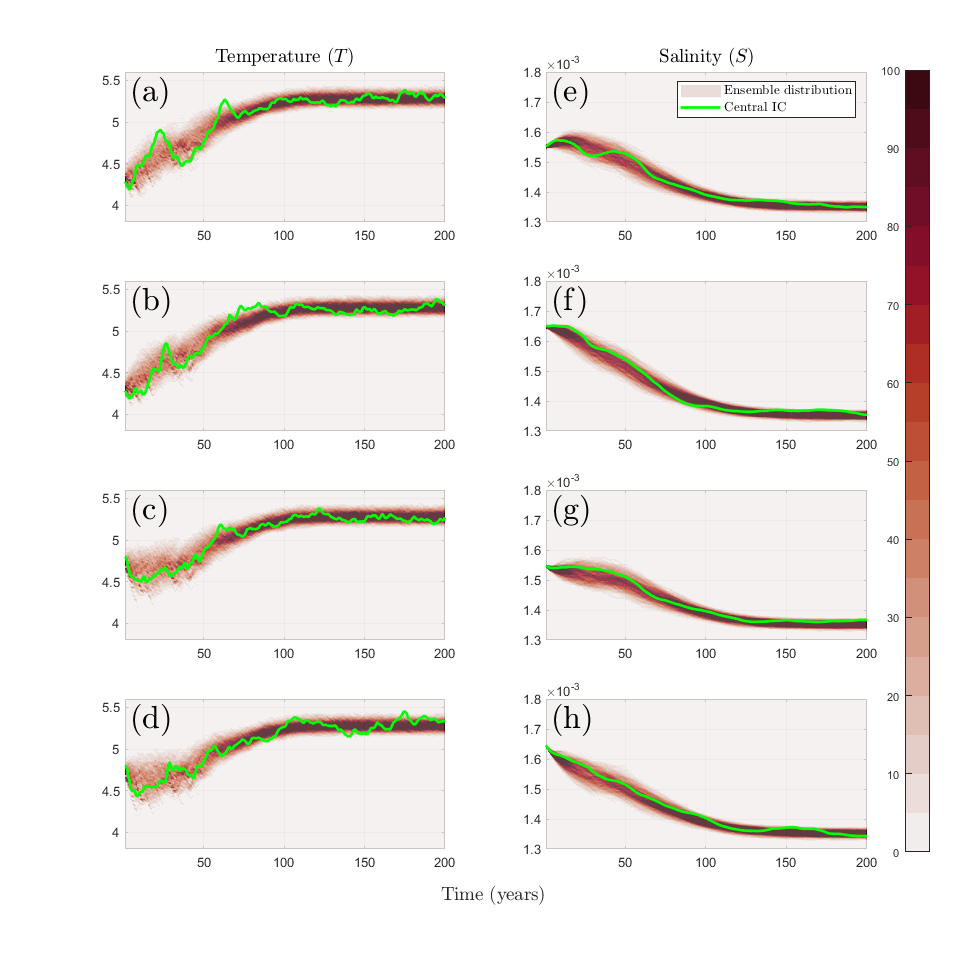}
\caption{Macro ICU in a single variable: comparing the evolution set and distribution of the slow-mixing ocean variables for different micro ICEs in a macro ICU scenario, with $H=0.01$ in the first 100 years, and $H=0$ in the remaining 100 years. Left column shows the ocean temperature. Right column shows ocean salinity. Panels (a-f) show: (a,e) IC 5; (b,f) IC 6; (c,g) IC 7; and (d,h) IC 8.}
\label{fig6}
\end{figure}

The sensitive to macro ICU with respect to a single variable is illustrated in Figure~\ref{fig6}, which shows the results for micro ICEs starting from the ICs indicated in Figure~\ref{fig5}. Note that macro uncertainty in salinity does not seen to alter the evolution set and its distribution, as indicated in Figure~\ref{fig6}(a,b). On the other hand, macro uncertainty in temperature has a significant effect on salinity, as shown in Figures~\ref{fig6}(e,g): the evolution set ans its distribution for salinity are significantly different, despite having ensembles around the same initial salinity state (see also Figure S.4).

This sensitivity of both $\mathcal{E}$ and $\mu_{\mathcal{E}}$ to macro ICU in a single, slow variable is remarkable, and suggests that a proper quantification of the uncertainty in future climate projections requires an assessment of macro ICU as well.

\subsection{Convergence time and macro ICU}\label{Sec5.3}

\begin{figure}
  \centering
\includegraphics[width=1\textwidth]{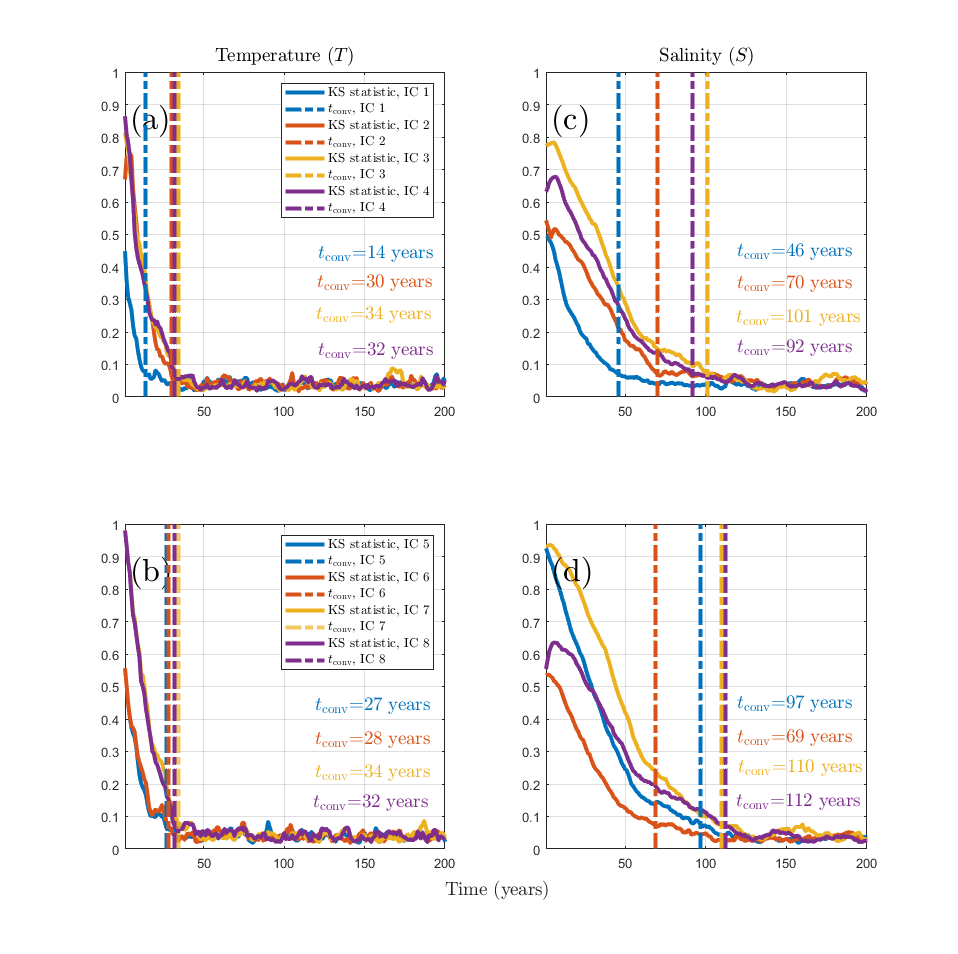}
\caption{Distance between the micro ICE distributions to the pullback invariant distribution, measured through the KS statistics (solid lines), and convergence time (dashed-dot lines) computed using Equation~(\ref{eq3.1-1}): (a,b) ocean temperature; (c,d) ocean salinity, for the micro ICEs centred at: ICs 1 to 4 (left column) as per Figure~\ref{fig4}; ICs 5 to 8, as per Figure~\ref{fig6}.}
\label{fig7}
\end{figure}

As macro ICU impact the evolution set and its distribution, one might asks whether the convergence time $t_{\mathrm{conv}}$ is also affected by it. Here we revisit the concept of convergence time and show how it can vary with in a macro ICU scenario. We illustrate this by computing $t_{\mathrm{conv}}$, using Equation~(\ref{eq3.1-1}), for the eight micro ICEs shown in Figure~\ref{fig3} and Figure~\ref{fig5}. The resulting $t_{\mathrm{conv}}$ and corresponding evolution of the KS statistics are shown in Figure~\ref{fig7} for the ocean variables. 

When starting from a control run trajectory (as per Figure~\ref{fig3}), the resulting $t_{\mathrm{conv}}$ can vary dramatically. This is shown in Figures~\ref{fig7}(a,c). First, IC 1 provides a short $t_{\mathrm{conv}}$ for both temperature and salinity, being of 14 years and 46 years, respectively.
This $t_{\mathrm{conv}}$ increases substantially from IC 1 to IC 2, being of 30 years for temperature and 70 for salinity. For IC 3 and IC 4, while the $t_{\mathrm{conv}}$ for temperature remains of the same order (34 and 32 years, respectively), it still varies substantially for salinity, resulting in a $t_{\mathrm{conv}}$ of 101 years for IC 3 and and 92 years for IC 4. We also note that the order of $t_{\mathrm{conv}}$ is the same for both variables in this case: IC 1 has shortest $t_{\mathrm{conv}}$ for both temperature and salinity, IC 2 is second, and so on.

When starting from chosen-values within the attractor (as per Figure~\ref{fig5}), the results are rather different. This is shown in Figures~\ref{fig7}(b,d). In particular, both variability and and order of $t_{\mathrm{conv}}$ differs from those shown in Figures~\ref{fig7}(a,c). For instance, the variability in $t_{\mathrm{conv}}$ is 27 to 34 years for temperature (instead of 14 to 34 years) and 69 to 112 years (instead of 46 to 112) for salinity. Also, the shortest $t_{\mathrm{conv}}$ for temperature (27 years) is given by IC 5, while the shortest for salinity is given by IC 6 (69 years).

The starkest contrast is observed when comparing IC 6 and IC 8. Note that, while both ICs have the same value of salinity, their respective micro ICEs have a $t_{\mathrm{conv}}$ that differs by 43 years, highlighting the impact that macro ICU in a single variable (in this case ocean temperature) can have in other variables. 

In Herein et al. (2016)~\cite{herein2016:snapshotGCM}, where the authors only looked at uncertainty using an intermediate-complexity model, they noted that $t_{\mathrm{conv}}$ did not chance for micro ICEs starting at different instants of time. However, their micro ICE was generated perturbing only one variable (the surface pressure field), keeping the others equal for all ensemble members, while the results correspond to another variable, the annual mean surface temperature in a single grid point of the model (what they called a small scale) located within continental Europe. While treated there as a simple approximation, the results presented here for a much simpler model suggests that uncertainty in other variables can have a significant impact on the distribution and its convergence time, as the response from the initial uncertainty in slow variables could take a while to show.

\section{How rate of change in forcing affects the uncertainty of climate predictions}\label{Sec6}

In the context of climate change, the system is under an external forcing (e.g. change in temperature due to anthopogenic carbon dioxide emissions) that is both dynamic and uncertain. Those uncertainties are usually investigated via scenarios, which in the context of IPCC, have shown to dramatically affect the climatology predicted by CMIP models. In the context of this work, this external forcing uncertainty may also affect the evolution of an ICE as a distribution. 

\begin{figure}
  \centering
\includegraphics[width=1\textwidth]{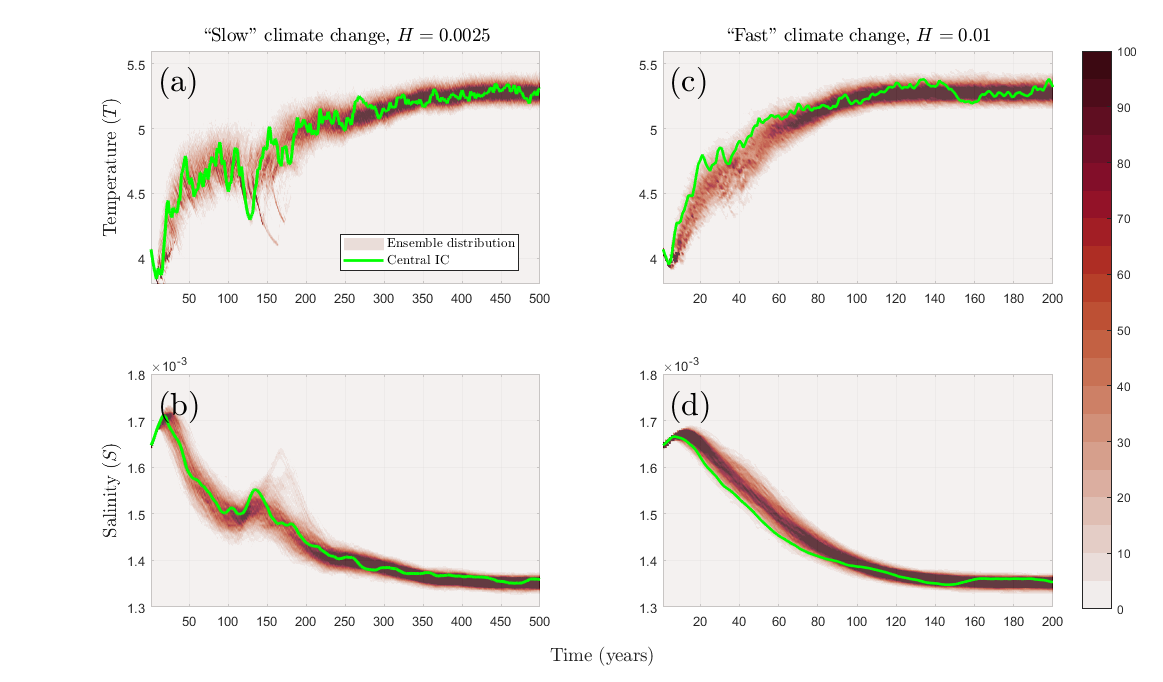}
\caption{ICE distributions starting from IC 2 in Figure 3, for $H=0.01$ (100 years of climate change, followed by 100 years of non-forced climate with $F_{m}=8$) shown in the upper panels, and $H=0.0025$ (400 years of climate change, followed by 100 years of non-forced climate with $F_{m}=8$) in the bottom panels. Left column panels show temperature for (a) $H=0.01$; (b) $H=0.0025$. Right column panels show salinity for (c) $H=0.01$; (d) $H=0.0025$.}
\label{fig8}
\end{figure}

We illustrate this by looking at the evolution of the micro ICE centred in IC 2 (shown in Figure~\ref{fig3}) but under a slower rate of ``climate'' change regime. Here, we reduce the rate of change in forcing $H$ by a quarter, from $H=0.01$ to $H=0.0025$, meaning that it now takes 400 years for the baseline forcing $F_{m}$ to increase by one unit. The resulting time series are shown in Figure~\ref{fig8} for ocean temperature and salinity, where we also included the $H=0.01$ time series for reference.

Changing, or in this case reducing, the speed of climate change has important effects on the resulting distributions. While the ICE distributions in Figure~\ref{fig8}(c) shows a mildly monotonic decrease, Figure~\ref{fig8}(a) shows that this behaviour, while kept, is much more pronounced under a weaker forcing. As this slowly changing distribution evolves, it again shows a distinct behaviour at around year 120: the distribution suddenly gets broader, with the temperature of several ensemble members decreasing sharply. About 40 to 50 years later, the ensemble narrows again and regain a shape akin to that of Figure~\ref{fig8}(c). These behaviour are mirrored by the salinity distribution, as shown in Figures~\ref{fig8}(b,d).

\begin{figure}
  \centering
\includegraphics[width=1\textwidth]{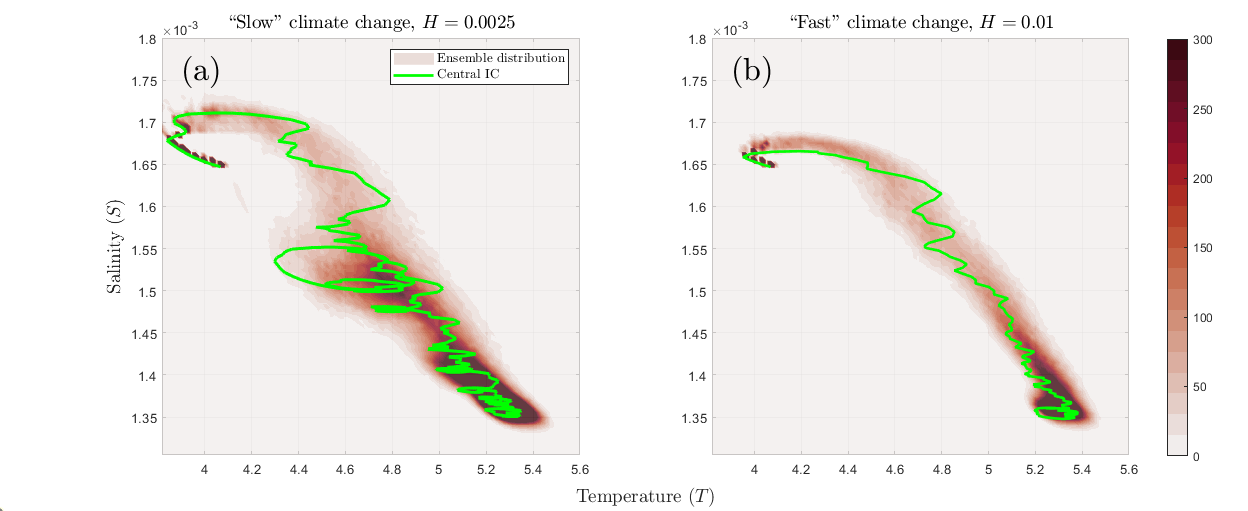}
\caption{Projection of the phase space onto the $(T,S)$ subspace, with a heatmap indicating the number of ensemble members that passes through each point at least once (no repetitions are counted): (a) $H=0.01$; (b) $H=0.0025$. These correspond to the joint distributions shown in Figure~\ref{fig8}(a,c) and Figure~\ref{fig8}(b,d), respectively.}
\label{fig9}
\end{figure}

\begin{figure}
  \centering
\includegraphics[width=1\textwidth]{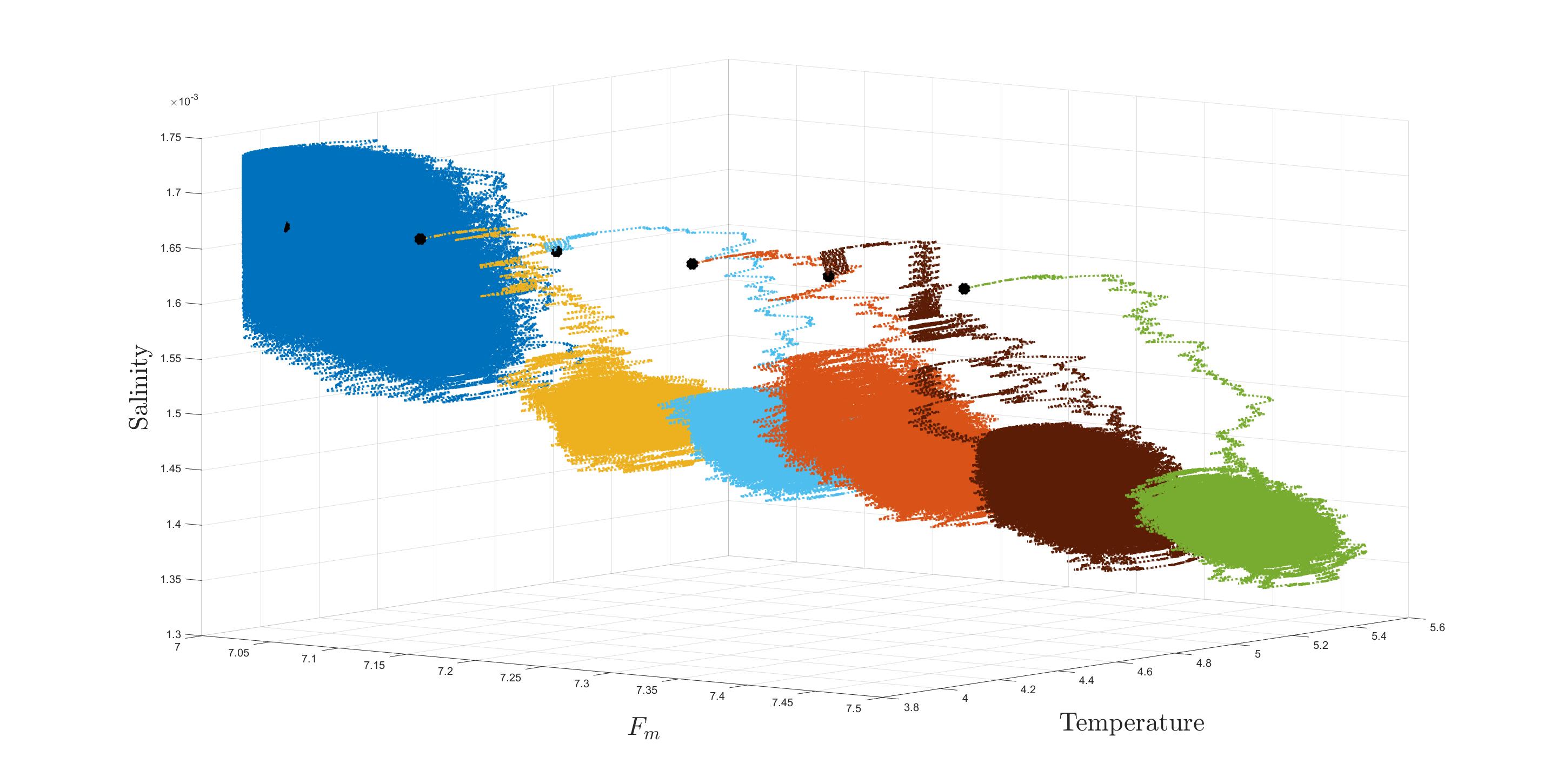}
\caption{Attractor for the non-forced L84-S61, projected over the ocean temperature-salinity $(T,S)$ subspace, for several values of $F_{m}$ between 7 and 7.5. All attractors shown correspond to a single trajectory starting from the same IC (black dots).}
\label{fig10}
\end{figure}

These curious behaviour, which is consequence solely of altering the rate of change in forcing, can be better seen when looking at the projection of the phase space onto the ocean variables subspace, as shown in Figure~\ref{fig9}. At a faster climate change rate, shown in Figure~\ref{fig9}(b), the distribution seems to have less freedom to explore the phase space and has its way forced into the attractor $F_{m}=8$. At a slower climate change rate, presented in Figure~\ref{fig9}(a), the ensemble members have now more freedom - and time - to explore the phase space and any intermediate attractors between those of $F_{m}=7$ and $F_{m}=8$. As suggested by Figure~\ref{fig10}, one of those intermediate attractors is somehow broader (in the ocean variables) than the neighbour ones, and trajectories entering there might eventually reach (time allowing) lower values of temperature and higher values of salinity.

\section{Conclusions}\label{Sec7}

This article discussed several aspects related to the climate predictability in short- and mid-time scales, including annual to multi-decadal. To do so, we introduced the idea of an evolution set, where we combined the concepts of pullback attractor and micro ICU to produce an object lying within the system's pullback attractor whose shape is constrained by a more refined knowledge of the initial state of the system - via a micro ICE. While the evolution set is usually contained in the pullback attractor set, the latter is much larger, and their respective distributions, or climate projections, are different.

In addition to that, we attempted at defining a convergence time, as the time taken for an ICE distribution to become indistinguishable from the pullback invariant distribution. We also explored micro and macro ICU, revisited the concept of pullback attractor, and discussed the influence of those in the evolution set and the convergence time. We also discussed the effect of different rates of change in forcing in the evolution set. Given the significant differences produced, these results suggest that all these aspects should be considered when designing ensembles for chaotic, non-autonomous systems, in particular for ESMs in a climate-change scenario - i.e. under non-periodic external forcing.

Although the results obtained are dependent on the particular low-dimensional model used, the ideas are model-independent and should be applicable to any chaotic non-autonomous system. This includes the concepts of evolution set, micro and macro ICU and convergence time.

From a theoretical point of view, this work leaves many questions to be answered, which we believe to be of both mathematical and climate science relevance. The first set of questions relate to the evolution set $\mathcal{E}=\mathcal{E}(t;t_{0},\mathbf{X}_{0},\sigma_{\mathbf{X}_{0}})$:

\begin{itemize}
 \item Is it possible to prove rigorous results regarding the sensitivity and dependence of $\mathcal{E}$ to the central IC $\mathbf{X}_{0}$, initial time $t_{0}$ and variance $\sigma_{\mathbf{X}_{0}}$?
 \item What is the relationship of $\mathcal{E}$ to the pullback attractor $\mathcal{A}$? Is there any other relationship beyond $\mathcal{E} \subseteq \mathcal{A}$ when $D_{\mathbf{X}_{0}} \subseteq \mathcal{A}(t_{0})$? 
 \item How many ensemble members are needed to characterise $\mathcal{E}$ for a given $\mathbf{X}_{0}$, $t_{0}$ and uncertainty as measured by $\sigma_{\mathbf{X}_{0}}$?
  \item How dependent is $\mathcal{E}$ on the shape of the ICE? For instance, would a non-Gaussian distribution lead to a very different $\mathcal{E}$?
 \item How does the distribution $\mu_{\mathcal{E}}$ relates to the pullback invariant distribution $\mu_{\mathcal{A}}$ of the pullback attractor?
\end{itemize}

Another important question is how does uncertainty in one variable propagates, or rather influence others? For example, we saw that macro ICU in temperature seems to greatly affect salinity, but the converse is not true.

A final but more ambitious question relates to the ``size'' of attractors, as illustrated in Figure~\ref{fig10}. How large are attractors in ESMs? In other words:
\begin{itemize}
\item Is it possible to estimate their shape without resourcing to brute force, given the computational limitations of running such models?
\end{itemize}

A final note on the evolution set $\mathcal{E}$ is that there can be many, depending on what observations ones uses to constrain the possible climate scenarios with. The same applies to the the evolution distribution $\mu_{\mathcal{E}}$. In this practical sense, the central IC and variance used in the definition of $\mathcal{E}$ are just fudges to simulate the residual uncertainty after the information from the observations has been brought in. So the questions above, although generically formulated, might be asked in relation to an $\mathcal{E}$ constructed from assimilating some observation into a more realistic climate model for example. Nevertheless, answers to those questions would be a valuable resource in the design of relevant and influential climate model ensembles.

\section*{Acknowledgements}
F.d.M.V. and D.A.S. acknowledge the support received by a UK Natural Environment Research Council grant (ODESSS, agreement number NE/V011790/1).

\section*{Data availability}
All data used in this work is freely available online on Zenodo~\cite{dmv2023:datasetChaos}.

\medskip

\bibliographystyle{unsrt} 
\bibliography{MAIN-2023deMeloVirissimoStainforthEvolutionSets}  
\end{document}


\begin{center}
\Large{\textbf{SUPPLEMENTARY MATERIALS}}\\
\end{center}
\begin{center}
\large{\textbf{THE EVOLUTION OF A NON-AUTONOMOUS CHAOTIC SYSTEM UNDER NON-PERIODIC FORCING: A CLIMATE CHANGE EXAMPLE}}\\
\end{center}

\begin{center}
F. DE MELO VIR\'ISSIMO$^{1,}$\footnote{Corresponding Author. e-mail: f.de-melo-virissimo@lse.ac.uk.}, D. A. STAINFORTH$^{1,2}$, J. BR\"{O}CKER$^{3}$\\
\end{center}

\begin{center}
{\small \it $^1$Grantham Research Institute on Climate Change and the Environment, London School of Economics and Political Science, London, WC2A 2AE, United Kingdom.}\\
{\small \it $^2$Department of Physics, University of Warwick, Coventry, UK}\\
{\small \it $^3$School of Mathematical Physical and Computational Sciences, University of Reading, Reading, RG6 6AX, United Kingdom}\\
\end{center}

\begin{center}
     \textbf{NOTE: THIS IS A NON-PEER REVIEWED PREPRINT SUBMITTED TO ARXIV}
\end{center}
\vspace{1cm}

\begin{figure}
  \centering
\includegraphics[width=1\textwidth]{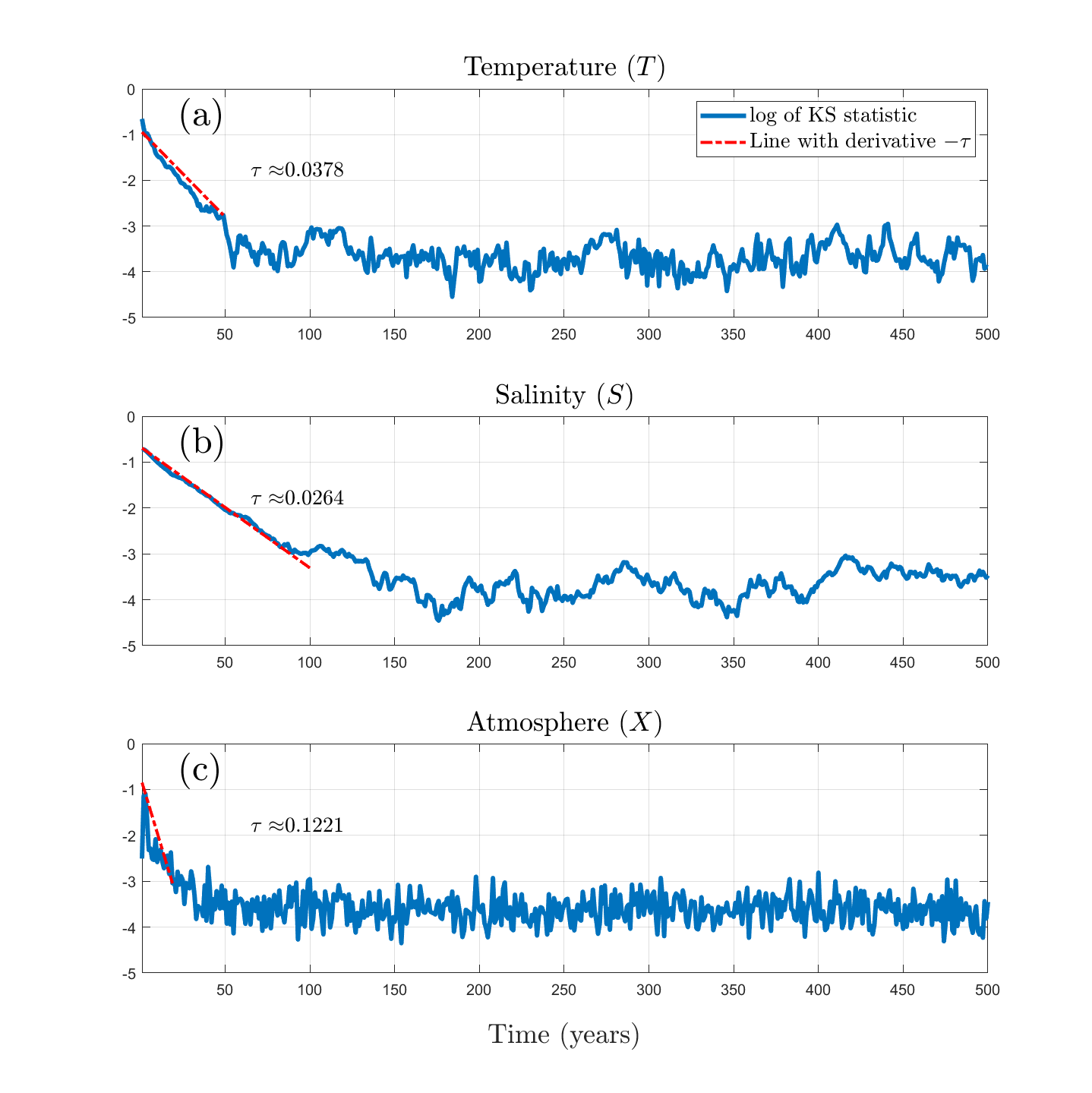}
\caption{Computing the decay time $\tau^{-1}$ for the distributions in Figure 1 (main manuscript) to loose most of its memory: (a) temperature; (b) salinity; (c) atmosphere $X$. To compute $\tau^{-1}$, one first computes the logarithm of the KS statistic (blue line) presented in Figures 1(d,e,f). Noting that the logarithm graph approximates a straight line in the first few years of decay, one can then estimate $-\tau$ as the angular coefficient (or derivative) or this straight line. To do so, we used data points at times: $t=3, 49$ years for temperature; $t=3,83$ years for salinity; $t=3,19$ years for the atmosphere variable $X$. This gives $\tau^{-1} \approx 26$ for temperature, $\tau^{-1} \approx 38$ for salinity, and $\tau^{-1} \approx 8$ for $X$. The red dot-dash line shows the corresponding line obtained from this reconstruction.}
\label{SM-fig1}
\end{figure}

\begin{figure}
  \centering
\includegraphics[width=1\textwidth]{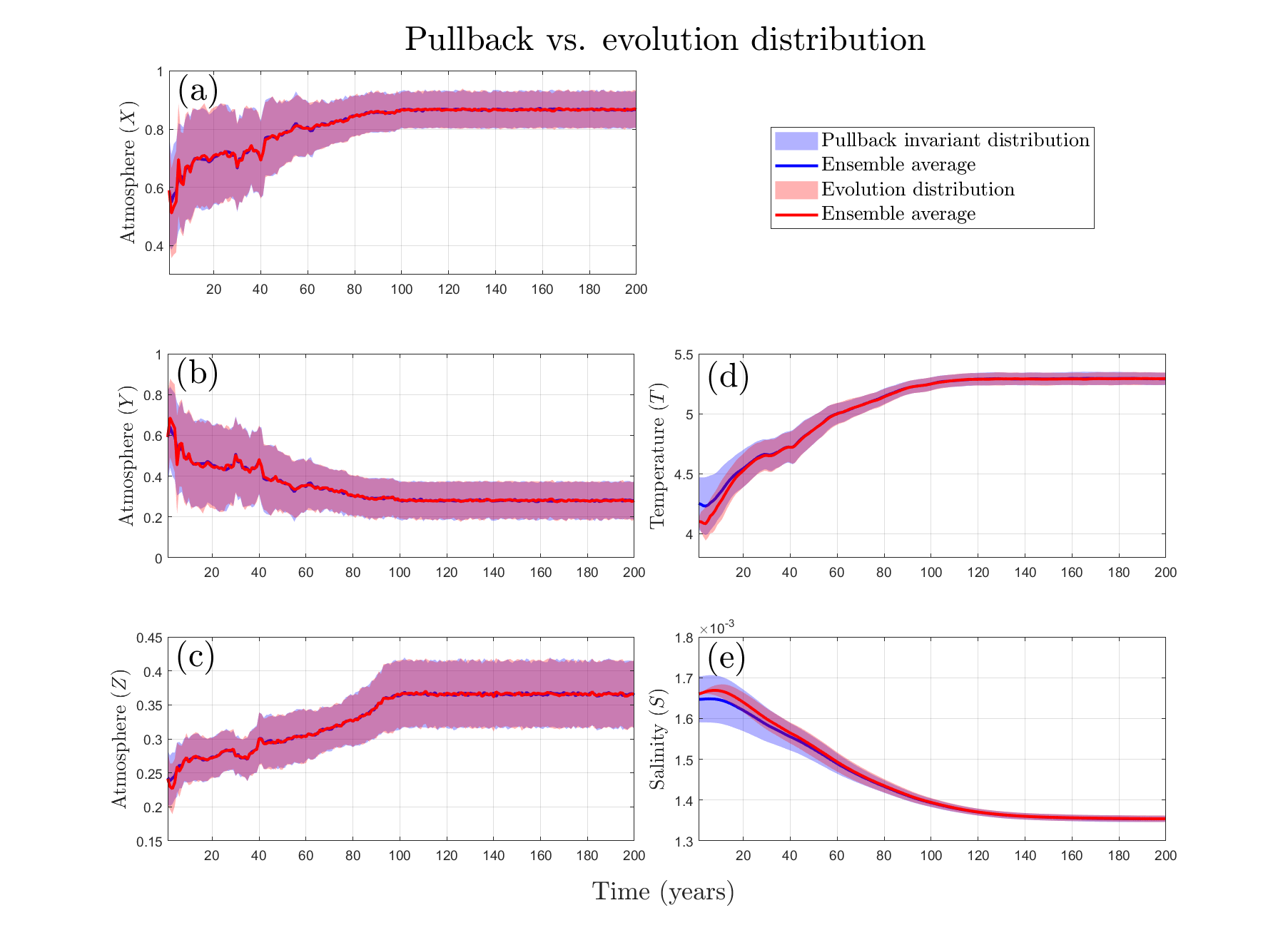}
\caption{Comparing the pullback invariant distribution with the evolution distribution generated by a micro ICE, with $H=0.01$ in the first 100 years, and $H=0$ in the remaining 100 years. Solid line shows the ensemble mean, and the shade shows 1 standard deviation from the mean. In blue is shown an ensemble that initially covers the entire pullback attractor. In red is shown the evolution of a micro ICE. Panels correspond to individual variables: (a) atmosphere $X$; (b) atmosphere $Y$; (c) atmosphere $Z$; (d) temperature; (e) salinity.}
\label{SM-fig2}
\end{figure}

\begin{figure}
  \centering
\includegraphics[width=1\textwidth]{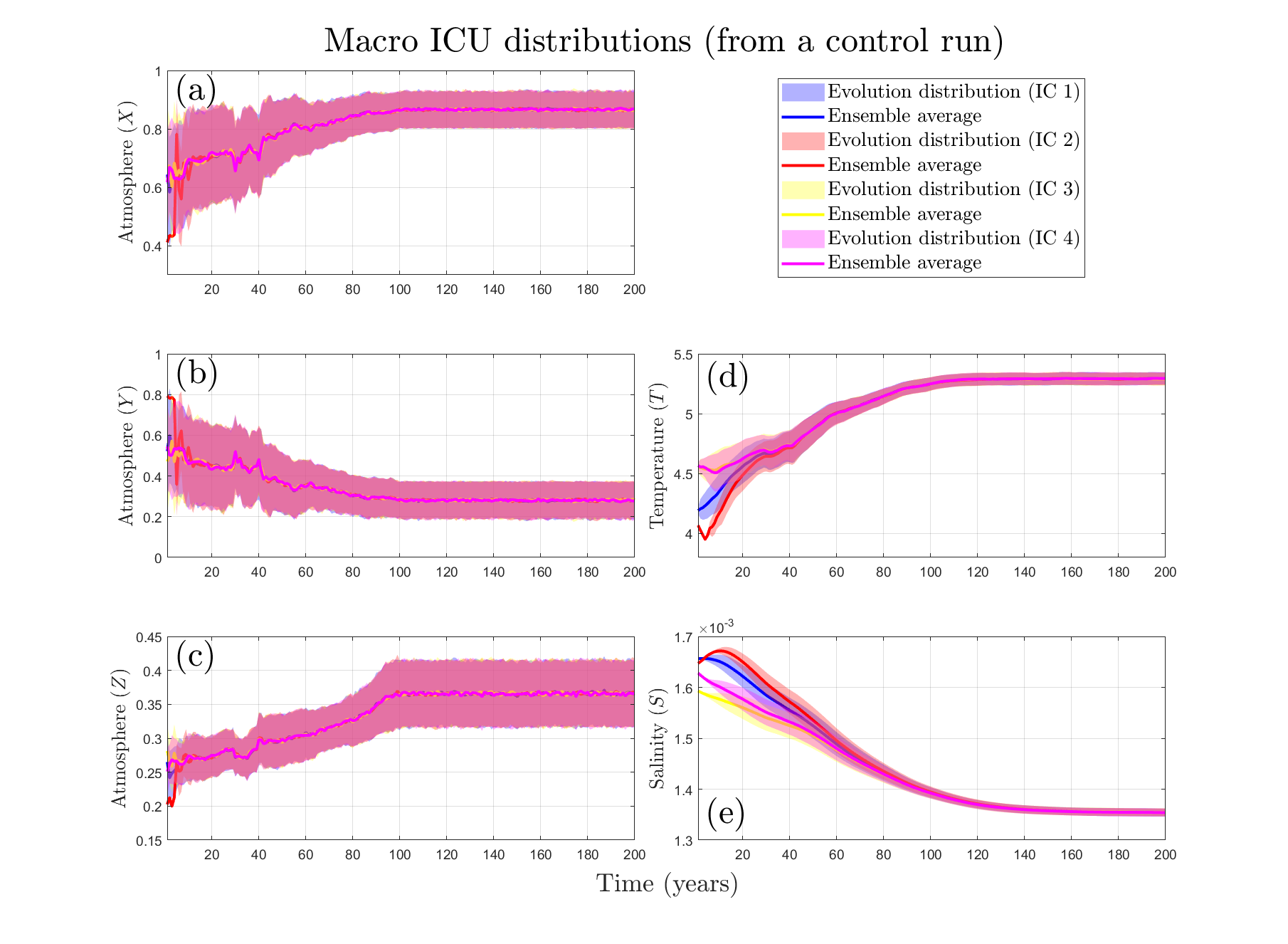}
\caption{Macro ICU from a control run simulation: comparing the evolution set and distribution of the slow-mixing ocean variables for different micro ICEs in a macro ICU scenario, with $H=0.01$ in the first 100 years, and $H=0$ in the remaining 100 years. Solid line shows the ensemble mean, and the shade shows 1 standard deviation from the mean. Results for IC 1, IC 2, IC 3 and IC 4 are presented in blue, red, yellow and magenta, respectively. Panels correspond to individual variables: (a) atmosphere $X$; (b) atmosphere $Y$; (c) atmosphere $Z$; (d) temperature; (e) salinity.}
\label{SM-fig3}
\end{figure}

\begin{figure}
  \centering
\includegraphics[width=1\textwidth]{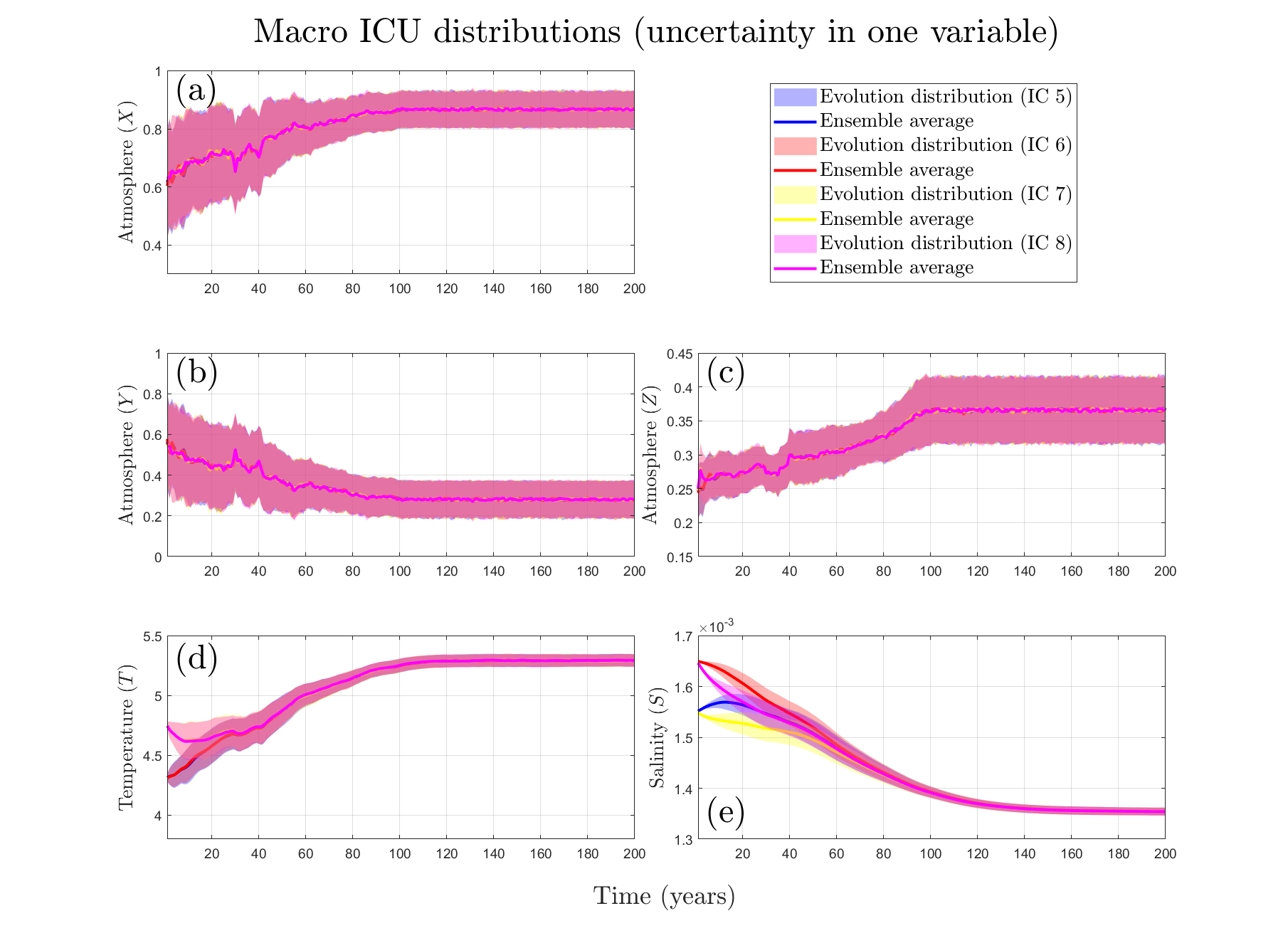}
\caption{Macro ICU in a single variable: comparing the evolution set and distribution of the slow-mixing ocean variables for different micro ICEs in a macro ICU scenario, with $H=0.01$ in the first 100 years, and $H=0$ in the remaining 100 years. Solid line shows the ensemble mean, and the shade shows 1 standard deviation from the mean. Results for IC 5, IC 6, IC 7 and IC 8 are presented in blue, red, yellow and magenta, respectively. Panels correspond to individual variables: (a) atmosphere $X$; (b) atmosphere $Y$; (c) atmosphere $Z$; (d) temperature; (e) salinity.}
\label{SM-fig4}
\end{figure}

{\small
\begin{center}
    \textbf{This PDF file includes:
    \begin{itemize}
        \item Supplementary text
        \item Figures~\ref{SM-fig1} to~\ref{SM-fig4}
        \item Table~\ref{SM-table1}
        \item SI References
    \end{itemize}
}
\end{center}
}

\newpage

\sloppy

\section{Model parameters}

The table below contains all model parameters and their values. The values shown were the same in all simulations.

\begin{table}[h!]
\centering
    \begin{tabular}{ c | c | p{10cm} }
    \hline
    \hline
    \textbf{Parameter} & \textbf{Value} & \textbf{Description} \\ \hline
    $F_{1}$ & 0.02 & Coupling parameter for the equator-pole temperature difference \\ 
    $G_{0}$ & 1 & Reference value for the land-sea temperature difference \\ 
    $G_{1}$ & 0.01 & Coupling parameter for the land-sea temperature difference \\
    $a$ & 0.25 & Damping coefficient of the westerly winds \\
    $b$ & 4 & Displacement of the waves due to interaction with the westerly wind\\
    $T_{av}$ & 30 & Standard temperature contrast between the polar and the equatorial box \\
    $\gamma$ & 30 & Proportionality constant between the westerly wind and non-homogeneous forcing by solar heating  \\
    $k_{w}$ & $1.8 \cdot 10^{-5}$ & Coefficient of internal diffusion in the ocean \\
    $k_{a}$ & $1.8 \cdot 10^{-4}$ & Coefficient of heat exchange between the ocean and atmosphere \\
    $\omega$ & $1.3 \cdot 10^{-4}$ & Coefficient derived from the linearised equation of state \\
    $\epsilon$ & $1.1 \cdot 10^{-3}$ & Coefficient derived from the linearised equation of state \\
    $\gamma_{0}$ & $7.8 \cdot 10^{-7}$ & Coefficient for the atmospheric water transport \\
    $\gamma_{1}$ & $9.6 \cdot 10^{-8}$ & Coupling parameter for the wind dependent atmospheric water transport \\ \hline \hline  
    \end{tabular}
  \caption{Description of the parameters and their reference values used in the L84-S61 model, as per Daron and Stainforth (2013)~\cite{daron2013:ensemble}}.
   \label{SM-table1}
\end{table}

\section{Experiments}

Here we describe in detail the experimental design of each simulation. The following holds for all experiments, unless otherwise noted:
\begin{itemize}
    \item {\bf Length of simulation:} 200 years, except experiments 1, 2, 3, 13 and 14.
    \item {\bf Time step:} 0.01 LTUs (1.2 hours).
    \item {\bf Output frequency:} 0.2 LTUs (1 day).
    \item {\bf Number of ensemble members:} 1,001 (1,000 plus central IC), except experiments 1 and 14.
    \item {\bf Variance:} $\sigma_{\mathbf{X}_{0}}=(0.02,0.02,0.02,0.002,0.000001)$, except experiments 1 and 14.
    \item {\bf Rate of change:} $H=0.01$ (1 unit every 100 years), except experiments 1, 2, 13 and 14.
    \item {\bf Year when climate change starts:} $(t_{\mathrm{start}}/K)=0$, except experiments 1, 2, 3 and 14.
    \item {\bf Year when climate change ends:} $(t_{\mathrm{end}}/K)=100$, except experiments 1, 2, 3, 13 and 14.
\end{itemize}

The ICs vary across experiments, and are presented in detail below. 

\subsection{Experiment 1: single trajectory, no climate change}

\begin{itemize}
    \item Initial condition: $\mathbf{X}_{0}=(1.012586,1.030767,-2.622185\times 10^{-1},4.080627,1.658932\times 10^{-3})$
    \item Length of simulation: 100,000 years
    \item Number of ensemble members: 1 (central IC only)
    \item Rate of change: $H=0$
\end{itemize}

\subsection{Experiment 2: Pullback attractor, no climate change}
\begin{itemize}
    \item Initial condition: $\mathbf{X}_{0}=(1.012586,1.030767,-2.622185\times 10^{-1},4.080627,1.658932\times 10^{-3})$
    \item Length of simulation: 500 years
    \item Rate of change: $H=0$
\end{itemize}

\subsection{Experiment 3: Pullback attractor, under climate change}

\begin{itemize}
    \item Initial condition: $\mathbf{X}_{0}=(1.012586,1.030767,-2.622185\times 10^{-1},4.080627,1.658932\times 10^{-3})$
    \item Length of simulation: 500 years
    \item Year when climate change starts: $(t_{\mathrm{start}}/K)=300$
    \item Year when climate change ends: $(t_{\mathrm{end}}/K)=400$
\end{itemize}

\subsection{Experiment 4: Micro ICE stating from pullback attractor, under climate change}

\begin{itemize}
    \item Initial condition: $\mathbf{X}_{0}=(1.012586,1.030767,-2.622185\times 10^{-1},4.080627,1.658932\times 10^{-3})$
\end{itemize}

\subsection{Experiments 5 to 8: Macro ICU starting from a spinup trajectory}

\noindent Experiment 5:
\begin{itemize}
    \item Initial condition: $\mathbf{X}_{0}=
(1.550668,-1.574188\times 10^{-1},1.380261,4.166778,1.656004\times 10^{-3})$
\end{itemize}

\noindent Experiment 6:
\begin{itemize}
    \item Initial condition: $\mathbf{X}_{0}=(4.617810\times 10^{-2},1.553979,-5.371011\times 10^{-2},4.084877,1.645639\times 10^{-3})$
\end{itemize}

\noindent Experiment 7:
\begin{itemize}
    \item Initial condition: $\mathbf{X}_{0}=(2.856310\times 10^{-1},1.218884,1.032842,4.537546,1.594370\times 10^{-3})$
\end{itemize}

\noindent Experiment 8:
\begin{itemize}
    \item Initial condition: $\mathbf{X}_{0}=(1.911878,1.623725,5.750266\times 10^{-1},4.564914,1.630071\times 10^{-3})$
\end{itemize}

\subsection{Experiments 9 to 12: Macro ICU starting from chosen values}

\noindent Experiment 9:
\begin{itemize}
    \item Initial condition: $\mathbf{X}_{0}=(1.012586,1.030767,-2.622185\times 10^{-1},4.3,1.55\times 10^{-3})$
\end{itemize}

\noindent Experiment 10:
\begin{itemize}
    \item Initial condition: $\mathbf{X}_{0}=(1.012586,1.030767,-2.622185\times 10^{-1},4.3,1.65\times 10^{-3})$
\end{itemize}

\noindent Experiment 11:
\begin{itemize}
    \item Initial condition: $\mathbf{X}_{0}=(1.012586,1.030767,-2.622185\times 10^{-1},4.75,1.55\times 10^{-3})$
\end{itemize}

\noindent Experiment 12:
\begin{itemize}
    \item Initial condition: $\mathbf{X}_{0}=(1.012586,1.030767,-2.622185\times 10^{-1},4.75,1.65\times 10^{-3})$
\end{itemize}

\subsection{Experiment 13: Micro ICE under slow climate change}

\begin{itemize}
    \item Initial condition: $\mathbf{X}_{0}=(4.617810\times 10^{-2},1.553979,-5.371011\times 10^{-2},4.084877,1.645639\times 10^{-3})$
    \item Length of simulation: 500 years
    \item Rate of change: $H=0.0025$ (1 unit every 400 years)
    \item Year when climate change starts: $(t_{\mathrm{start}}/K)=0$
    \item Year when climate change ends: $(t_{\mathrm{end}}/K)=400$
\end{itemize}

\subsection{Experiment 14: Attractors for different $F_{m}$ when $H=0$}

\begin{itemize}
    \item Initial condition: $\mathbf{X}_{0}=(1.012586,1.030767,-2.622185\times 10^{-1},4.080627,1.658932\times 10^{-3})$
    \item Length of simulation: 50,000 years
    \item Number of ensemble members: 1 (central IC only)
    \item Rate of change: $H=0$
    \item Note: there are six experiments, one for each value of $F_{m}=7,7.1,7.2,7.3,7.4,7.5$
\end{itemize}


\bibliographystyle{unsrt} 
\bibliography{SUPPLEMENT-2023deMeloVirissimoStainforthEvolutionSets}  
